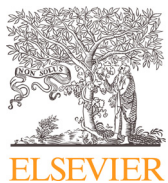
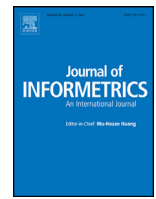

# Gender-based homophily in research: A large-scale study of man-woman collaboration

Marek Kwiek[a], Wojciech Roszka[b]

[a] *Institute for Advanced Studies in Social Sciences and Humanities (IAS), UNESCO Chair in Institutional Research and Higher Education Policy, Adam Mickiewicz University in Poznan, Poland*
[b] *Poznan University of Economics and Business, Poznan, Poland*

**ABSTRACT**

We examined the male-female collaboration practices of all internationally visible Polish university professors (N = 25,463) based on their Scopus-indexed publications from 2009–2018 (158,743 journal articles). We merged a national registry of 99,935 scientists (with full administrative and biographical data) with the Scopus publication database, using probabilistic and deterministic record linkage. Our unique biographical, administrative, publication, and citation database ("The Polish Science Observatory") included all professors with at least a doctoral degree employed in 85 research-involved universities. We determined what we term an "individual publication portfolio" for every professor, and we examined the respective impacts of biological age, academic position, academic discipline, average journal prestige, and type of institution on the same-sex collaboration ratio. The gender homophily principle (publishing predominantly with scientists of the same sex) was found to apply to male scientists—but not to females. The majority of male scientists collaborate solely with males; most female scientists, in contrast, do not collaborate with females at all. Across all age groups studied, all-female collaboration is marginal, while all-male collaboration is pervasive. Gender homophily in research-intensive institutions proved stronger for males than for females. Finally, we used a multi-dimensional fractional logit regression model to estimate the impact of gender and other individual-level and institutional-level independent variables on gender homophily in research collaboration.

## 1. Introduction

Science is a collaborative enterprise, with (male and female) scientists collaborating internationally, nationally, and institutionally (Wuchty, Jones, & Uzzi, 2007; Wagner, 2018). However, this is *not* our topic: our focus is on male–male, female–female, and male–female (or mixed-sex) research collaboration rather than collaboration across countries and institutions. The dominating view in literature is that, on average, males collaborate more often with males, and females collaborate more often with females (Jadidi, Karimi, Lietz, & Wagner, 2018; Lerchenmueller, Hoisl, & Schmallenbach, 2019; Wang, Lee, West, Bergstrom, & Erosheva, 2019; Holman & Morandin, 2019; Boschini & Sjögren, 2007; McDowell & Smith, 1992). This hypothesis is being tested using a large-scale dataset with unique variables.

According to the homophily principle, "similarity breeds connection"; consequently, personal networks are homogeneous with regard to many sociodemographic and personal characteristics (such as age, ethnic origin, class origin, wealth, education, and gender). On the positive side, homophily is reported to simplify communication (McPherson, Smith-Lovin, & Cook, 2001; Kegen, 2013). However, on the negative side, homophily may "limit people's social worlds in a way that has powerful implications for the information

---

they receive, the attitudes they form, and the interactions they experience" (McPherson et al., 2001). As science is increasingly collaborative, the homophily principle may increasingly influence academic careers. Research collaboration in science (or gender co-authorship patterns) provides fertile ground to test the homophily principle.

Man–woman research collaboration patterns in science are contrasted in this paper through six lenses: biological age, academic position, academic discipline, gender-defined research collaboration type, journal prestige, and institutional research intensity. The individual scientist, rather than the individual article, is the unit of analysis. The key innovative methodological step is the determination of what we term an "individual publication portfolio" (for the decade of 2009–2018) for every internationally visible Polish scientist (N = 25,463 university professors from 85 universities, grouped into 27 disciplines, along with their 164,908 international collaborators, who together authored 158,743 Scopus-indexed publications). Co-authorships are used for the operationalization of research collaboration, following standard bibliometric practice.

The individual publication portfolio reflects the distribution of gender-defined research collaboration types (same-sex collaboration and mixed-sex collaboration) for every individual scientist. Team formation in academia, understood as publishing with coauthors of varying numbers and different genders, is voluntary (McDowell & Smith, 1992): researchers team up when they think that they are better off collaborating than publishing alone. The teams formed, or the articles published, are likely to reflect "individual tastes and perceptions of the returns to collaboration, as well as the costs of coordination" (Boschini & Sjögren, 2007, p. 327). Some male scientists collaborate predominantly with other males, and some female scientists collaborate predominantly with other females. Still, others prefer to publish in mixed-sex collaborations (or to author individually). We examine the same-sex collaboration ratio at an individual level of every internationally visible Polish scientist (i.e., only authors with Scopus-indexed publications) and generalize the results from the individual level to the level of the national higher education system.

## 2. Literature review

### 2.1. The gender context of science

The gender context of academic science has changed substantially in the past few decades (Huang, Gates, Sinatra, & Barabàsi, 2020; Larivière, Ni, Gingras, Cronin, & Sugimoto, 2013), with more female scientists entering the higher education sector (Elsevier, 2018) and occupying high academic positions (Zippel, 2017; Diezmann & Grieshaber, 2019). Male and female scientists often pursued or were pushed onto somewhat different career tracks and were located in different academic structures, with "differential access to valuable resources" (Xie & Shauman, 2003, p. 193). Females, as new entrants into a traditionally male-dominated academic profession, initially did not have equal access to professional networks (McDowell, Singell, & Stater, 2006). But the academic world is changing.

New bibliometric literatures applying the various gender-determination methods to authors and authorships (Halevi, 2019; Elsevier, 2020) bring new data-driven insights to gender disparities in science, and literatures have become much less based on anecdotal and localized studies (Larivière et al., 2013). Women are plugging into networks over time as the profession becomes more gender representative (as shown for academic economists by McDowell et al., 2006, p. 154). However, somewhat paradoxically, the increased participation of women in STEM disciplines is reported to have been accompanied by an increase in gender differences regarding both productivity and impact (Huang et al., 2020, p. 8; Elsevier, 2018, p. 16).

As recent literature highlights, female scientists occupy more junior positions and receive lower salaries, are more often in non-tenure-track and teaching-only positions, are promoted more slowly, are less likely to be listed as either first or last author on a paper, and are allocated less research funding from national research councils. Women also tend to be less involved in international collaboration; female collaborations are more domestically oriented than are the collaborations of males from the same country; and females have less-prestigious collaborations and fewer collaborations overall (see Holman & Morandin, 2019; Halevi, 2019; Larivière et al., 2013; Larivière et al., 2011; Aksnes, Rørstad, Piro, & Sivertsen, 2011; Aksnes, Piro, & Rørstad, 2019; Huang et al., 2020; Maddi, Larivière, & Gingras, 2019; Fell & König, 2016; van den Besselaar & Sandström, 2016; Nielsen, 2016). In every country studied recently in Elsevier (2020) and Elsevier (2018), the percentage of women who publish internationally is lower than the percentage of men who do so; for Poland, which is not included in the Elsevier reports, these publishing patterns are confirmed for various collaboration intensity levels and for various age groups; see Kwiek & Roszka, 2020, on gender disparities in international collaboration).

Female scientists in Poland constitute a substantial, highly productive, and highly internationalized part of the academic workforce, which is often the case in formerly communist European countries, which exhibit greater gender parity than the world and the OECD averages (Larivière et al., 2013, p. 212). Poland has a higher proportion of professors than any country studied in Larivière et al. (2013) or in Diezmann and Grieshaber (2019), reaching 29.82% in 2018 (GUS, 2019, p. 220), even though there is a clear "the higher the fewer" pattern across all institutional types. As will be highlighted by the results of this paper, in Poland, males tend to collaborate with males—but females tend *not* to collaborate with females. Thus, gender homophily is high among Polish males and low among females, the latter constituting 41.5% of Polish university professors (of all ranks, all with at least a doctoral degree) in our sample and 46.10% of the entire full-time academic workforce in 2018 (GUS, 2019, p. 220).

A recent cohort analysis of the effects of gender on the publication patterns in mathematics (Mihaljević-Brandt, Santamaria, & Tullney, 2016, pp. 8–13), one of the most heavily male-dominated academic fields, based on the scholarly output of 150,000 mathematicians, shows that women publish less at the beginning of their careers; they leave academia at a higher rate than men; and high-ranked mathematics journals publish fewer articles authored by women. Women may also suffer from "biased attention" to their work, even if their work is of comparable quality (Lerchenmueller et al., 2019, p. 10). The authors' gender is also reported to affect the citations received (Potthoff & Zimmermann, 2017; Maddi et al., 2019). Female lead authors are reported to receive up to 29% fewer citations for work published in the most influential journals (as shown for publications from the PubMed database of 3,233 recipients of prestigious fellowships in life sciences in the U.S.: Lerchenmueller et al., 2019, p. 4).

Furthermore, gender-based homophily in citations exists in all disciplines, as a study of the citation data of seven million articles published in 2008–2016 shows: the citer disproportionately cites references from authors who are of the same gender, male scientists disproportionately citing other male scientists, possibly leading to a "perpetual disparity" in citations in favor of men as men represent about 70% of all authorships (Ghiasi, Mongeon, Sugimoto, & Larivière, 2018, p. 1520). Moreover, recent research based on a sample of CVs of U.S. economists reports that gender influences the attribution of credit for group work, that is, co-authorship matters differently for tenure for men and women, with women being less likely to receive tenure the more they co-author (Sarsons, Gërxhani, Reuben, & Schram, 2021). This differential attribution of credit contributes to the gender promotion gap (Fell & König, 2016; Abramo, D'Angelo, & Rosati, 2015). Furthermore, the gender citation gap persists: even though female scientists may publish more in journals with higher impact factors than their male peers, their work may receive lower recognition (fewer citations) from the scientific community (as Ghiasi, Larivière, & Sugimoto, 2015, have shown for female engineers, using a sample of 680,000 articles from 2008–2013, and Maliniak et al., 2013, for top journals in international relations).

*2.2. Female scientists and competition*

Of the various approaches to studying the "increasing and persistent" gender gap (Huang et al., 2020, p. 3) and "pervasive" gender hierarchies (Fox, 2020, p. 1001) in science, an approach centered on competition is especially relevant in the context of homophilous and heterophilous collaboration patterns. There have been ongoing discussions in experimental and personnel economics (often with laboratory-based evidence) about whether women are deterred by competition in some areas of science (and in some workplaces more generally; Flory et al., 2014; Dargnies, 2012). The systematic shying away from competition could have implications not only for the gender distribution of females across academic disciplines and their sub-disciplines but also for team formation in research collaboration, selected prestige level of journals in academic publishing, and authorship composition. Laboratory experiments show that women may shy away from competition and men may embrace it, with gender implications for publishing in top academic journals, where competition is stiff and the risk of rejection high (Sonnert & Holton, 1996; Kwiek, 2021). Women are extremely underrepresented in top journals in some disciplines, such as mathematics (Mihaljević-Brandt et al., 2016, p. 19), and they can self-select into lower-ranked journals (Mayer & Rathmann, 2018). Gender differences in the propensity to choose competitive environments (in our case, highly selective journals) are reported to be driven by gender differences in confidence and preferences for entering and performing in a competition (Niederle & Vesterlund, 2007, pp. 1098–1100). In their study of all full professors in psychology in Germany, Mayer & Rathmann (2018, pp. 1674-1676) show that in top journal publications, there are considerably more men with a high publication output, as well as considerably less men with a low publication output. Gender differences in choices over competition may be driven partly by men preferring competitive to non-competitive settings and by a significantly stronger aversion to competitive workplaces among women compared to men (Flory et al., 2014). Not surprisingly, male scientists over-cite (King et al., 2017; Maliniak et al., 2013), are better represented in top journals, and have higher visibility in science (Maddi et al., 2019).

Academic norms or expectations of conventional behavior may also matter: there may be a common social practice, particularly in male-dominated disciplines of science, that "holds women up to more scrutiny than men" (Gupta, Poulsen, & Villeval, 2013, p. 16). Sonnert and Holton (1996, p. 69), in their study of gender disparities in career patterns of especially promising scientists, conclude that women might be seen as socialized to be less competitive "so that they choose their own niche rather than enter the fray with numerous competitors working on the same topic," often feeling they are "under the magnifying glass." Male scientists may be "more aggressive, combative and self-promoting in their pursuit of career success, and so they achieve higher visibility" (Sonnert & Holton, 1996, p. 67). Social norms may thus influence publishing patterns, including, for instance, predominantly same-sex publishing for male scientists—especially in more traditional societies such as Poland.

At the same time, in more firmly male-dominated disciplines (such as physics and astronomy, engineering, and computer science, in the Polish case), female scientists may feel more intense performance pressure due to their high visibility among the overwhelming majority of male scientists and carrying the burden of representing women in these disciplines. They may have to work "twice as hard to prove their competence," with all their actions being public, as Kanter (1977, p. 973) suggested in her classic study of the role of male-female proportions in workplace settings. Being less competitively inclined in an increasingly competitive environment of global science may hurt female scientists, especially in their early careers, at an individual level of obtaining tenure, salary increases, and research funding (Van den Besselaar & Sandström, 2015; Sarsons et al., 2021; Kwiek, 2018a). In Polish academia, the list of disciplines where female participation is approximately or exceeds 50% goes beyond the social sciences and humanities (to include also business, economics, and econometrics; agricultural and biological sciences; medicine; chemistry; biochemistry, genetics and molecular biology; and psychology; see Table 16 in Data Appendices). Out of the 24 ASJC Scopus disciplines studied in this paper, female representation reaches at least 50% in 10 of them.

*2.3. Gender homophily in research collaboration defined*

The literature investigating gender homophily in academic publishing is based both on research on selected institutions (e.g., McDowell & Smith, 1992), selected disciplines (predominantly economics, as in Boschini & Sjörgen, 2007, or McDowell, Singell, & Stater, 2006), and large-scale bibliometric data (see Wang et al., 2019, who examined 252,413 papers with 807,588 authorships from the JSTOR corpus, or Ghiasi et al., 2015, who studied approximately one million Web of Science authorships in engineering).

Most recent bibliometric studies on gender differences in research collaboration patterns suggest that men tend to co-author with men and women with women—leading to the research theme of "gender homophily" in science (Ghiasi et al., 2018; Potthoff & Zimmermann, 2017; Lerchenmueller et al., 2019; Kegen, 2013; Wang et al., 2019; Boschini & Sjögren, 2007). At the same time, however, collaboration in research, traditionally operationalized as co-authored publications, influences career progress. Excessive gender homophily among women, while supportive for early-career female researchers, may also harm their careers. This is especially relevant for particularly able female scientists publishing in high-impact journals (as Lerchenmueller et al., 2019, show with powerful empirical evidence). Women may place themselves at a disadvantage when collaborating disproportionately with other women because, for example, "women tend to be part of less resource-rich and influential networks or because women's work may receive less attention than men's, likely harming career progress" (Lerchenmueller et al., 2019, p. 3). This is not the case in Poland, though, as we shall demonstrate, since the Polish female scientists studied tend to avoid publishing exclusively with other female scientists at all levels of their careers and for all age groups.

As mentioned, the homophily principle maintains that "similarity breeds connection" and personal networks are homogeneous with regard to sociodemographic, behavioral, and intrapersonal characteristics. Homophily is known to "limit people's social worlds" (McPherson, Smith-Lovin, & Cook, 2001, p. 415). According to this principle, contact between similar people occurs at a higher rate than among dissimilar people; in other words, "birds of a feather flock together" (McPherson et al., 2001, p. 417). Thus, males should co-author with males in a disproportionate fashion, while females should co-author disproportionately with females, across countries, disciplines, and institutions.

Homophily, in general, (including the gender-based homophily examined in this research) is reported to simplify communication, enhance the predictability of behavior, entail reciprocity in collaboration, and increase trust between collaborating parties (McPherson et al., 2001, p. 435; Kegen, 2013, p. 63). As Kegen (2013, p. 65) notes, while the behavior of collaborators might be more predictable and collaboration potentially less costly, gender homophily might also exclude women from informal networks. Furthermore, embeddedness in academic social networks—especially informal networks—is crucial both for doing research and for achieving a career. "Networks matter. Producing high-quality work is not sufficient for research to gain the attention of the widest number of scholars or have the greatest impact" (Maliniak et al., 2013, p. 918).

If homophily means "the tendency of people to choose to interact with similar others" (McPherson et al., 2001, p. 435), then gender-based homophily in this research means Polish male scientists disproportionately co-authoring with other male scientists, and Polish female scientists co-authoring disproportionately with other female scientists. Recent research tends to indicate that female scientists exhibit stronger gender homophily than male scientists (Jadidi et al., 2018): females are reported to collaborate more often with females than males with males (Kegen, 2013; Lerchenmueller et al., 2019; Ghiasi et al., 2018). Evidence from co-authorship patterns in economics indicates that team formation in academic publishing is not gender-neutral: rather, there is powerful gender sorting in team formation (Boschini & Sjögren, 2007). However, the practices of collaboration between males and females differ across disciplines (Maddi et al., 2019); the patterns of international research collaboration differ cross-nationally (see Kwiek, 2020a, on 28 European countries) and between genders intra-nationally (see Kwiek, 2020b, and Kwiek & Roszka, 2020, on Poland).

*2.4. Hypotheses of this research*

Following a comprehensive literature review and based on prior in-depth knowledge of the Polish academic science system, we have formulated the following seven research questions leading to seven hypotheses (which are presented in Table 1, along with the results of our research):

The Polish science and higher education systems have been studied intensively. For instance, Kulczycki and colleagues examined the funding system (Kulczycki, Korzeń, & Korytkowski, 2017), Bieliński and Tomczyńska (2018) studied the various manifestations of the ethos of science and showed how Poland is moving away from Michael Polanyi's "republic of science". Feldy and Kowalczyk (2020) studied how scientists view the system of financing science, and Kulczycki and Korytkowski (2020) examined changing publication patterns in Poland. Furthermore, higher education reforms (e.g., Shaw, 2019; Antonowicz, Kulczycki, & Budzanowska, 2020; Kwiek, 2012), international research collaboration (Kwiek, 2020b), and high research productivity (Kwiek, 2018b) have been examined. Gender disparity in Polish science, however, has rarely been studied, and gender collaboration patterns, including gender homophily, have not been examined except by Kwiek and Roszka (2020), who studied international research collaboration by gender and showed that male scientists dominate in this collaboration type at each level of intensity, with significant cross-disciplinary differences (Nielsen, 2016, came to similar conclusions in his study of a Danish university). Siemienska (2007) examined gender research productivity gap referring to cultural capital of faculty members. Finally, Kosmulski (2015) analyzed the productivity and impact of male and female scientists in the period 1975–2014, based on a limited set of authors bearing one of the 26 most popular "–ski" or "–cki" names, showing that male scientists generally have higher productivity and impact than female scientists, except for in biochemistry, where their productivity and impact are almost equal.

## 3. Data and methods

*3.1. Dataset*

Two large databases of different natures were merged: Database I was an official national administrative and biographical register of all Polish academic scientists; Database II was the Scopus database. The two were merged to create "The Polish Science Observatory," which was maintained and periodically updated by the two authors (a short description of the database is presented in Kwiek and Roszka, 2020). The main steps in merging the biographical and administrative dataset (The Polish Science) with the publication and citation database (Scopus) are graphically shown in Fig. 1.

**Table 1**
Research hypotheses and results (summary).

| Research Question | Hypothesis | Result |
| --- | --- | --- |
| RQ1. What is the relationship between gender and same-sex collaboration? | Hypothesis 1. We would expect that the same-sex collaboration ratio is higher for female than for male scientists. | Not confirmed |
| RQ2. What is the relationship between gender, same-sex collaboration, and age? | Hypothesis 2. We would expect that the same-sex collaboration ratio decreases with age for both male and female scientists. | Confirmed for male scientists only |
| RQ3. What is the relationship between gender, same-sex collaboration, and academic position? | Hypothesis 3. We would anticipate that the same-sex collaboration ratio decreases with academic position for both male and female scientists. | Confirmed for male scientists only |
| RQ4. What is the relationship between gender, same-sex collaboration, and academic disciplines? | Hypothesis 4. We would anticipate that the same-sex collaboration ratio is higher in male-dominated academic disciplines. | Confirmed |
| RQ5. What is the relationship between gender, same-sex collaboration, and institutional research intensity? | Hypothesis 5. We would expect that the same-sex collaboration ratio is higher in research-intensive universities. | Confirmed for male scientists only |
| RQ6. What is the relationship between gender, gender-defined research collaboration type, and journal prestige? | Hypothesis 6. We would expect that the journal prestige level of mixed-sex publications is higher than that of same-sex publications for both male and female scientists. | Confirmed |
| RQ7. What is the impact of gender and other individual-level and institutional-level independent variables on gender homophily in research collaboration? | Hypothesis 7. In a fractional logit regression model, we would anticipate that individual-level independent variables are more influential than institutional-level independent variables in predicting the same-sex collaboration ratio. | Not confirmed |

**Table 2**
An example of probabilistic integration output (identical, similar, and disparate pairs of strings).

| Last name, Database II | First name, Database II | Last name, Database I | First name, Database I | Last name compliance | First name compliance | Posterior probability |
| --- | --- | --- | --- | --- | --- | --- |
| Kwiek | Marek | Kwiek | Marek | 2 | 2 | 0.9975556 |
| Mrowiec | Bozena | Mrowiec | Bożena | 2 | 1 | 0.9946168 |
| Sobkow | Agata | Sobków | Agata | 1 | 2 | 0.9991700 |
| Wltek | Bozena | Witek | Bożena | 1 | 1 | 0.9073788 |
| Mudry | Z. | Mudryk | Zbigniew | 2 | 0 | 0.8846165 |

Database I (created by the OPI National Research Institute) comprised 99,535 scientists employed in the Polish science sector as of November 21, 2017. Only scientists with at least a doctoral degree (70,272) and employed in the higher education sector were selected for further analysis (54,448 or 54.70% of all scientists, all working at 85 universities of various types). The data used were both demographic (gender and date of birth) and professional (the highest degree awarded; award date of Ph.D., habilitation, and full professorship; and institutional affiliation), with each scientist identified by a unique ID. Database II included 169,775 names from 85 institutions whose publications for the decade analyzed (2009–2018) were included in the database and 384,736 Scopus-indexed publications. Authors in Database II were defined by their institutional affiliations, Scopus documents, and individual Scopus IDs. Scopus uses a sophisticated author-matching algorithm to precisely identify publications by the same author; gender is not captured in Scopus Author Profiles (Elsevier, 2020, p. 119). We did not reconstruct the full publishing careers (as in Huang et al., 2020) of Polish scientists but only for the last decade, when their Scopus publications increased markedly.

We have identified authors with their different individual IDs in the two databases and provided them with a new ID in the new "Observatory" database. Probabilistic methods of data integration were used (Fellegi & Sunter, 1969; Herzog, Scheuren, & Winkler, 2007; Enamorado, Fifield, & Imai, 2019). Separately within each of the 85 universities, the first name and last name records of each record in Database I were compared with each of the records in Database II using the Jaro-Winkler string distance (with values from 0 to 1; see Jaro, 1989; Winkler, 1990). Pairs of strings with a distance greater than 0.94 were considered identical (signified by 2) (see Table 2). Pairs with a distance greater than 0.88 but less than 0.94 were considered similar (signified by 1), while those with a distance less than 0.88 were considered disparate (signified by 0). Next, using an expectation maximization algorithm (Enamorado et al., 2019), the posterior probability that a given pair of records belongs to the same unit was estimated. If the probability was greater than 0.85, the pair was considered to be part of the same unit (as suggested by Harron et al., 2017). The computation was made using the fastLink R package (version 0.6.0).

By employing a probabilistic approach to the merging of the data sets, it was possible to estimate the uncertainty of the process and thus assess the quality of the new integrated database by calculating the percentage of records incorrectly classified as matches (false discovery rate, FDR) and as non-matches (false negative rate, FNR). Deduplication procedures were applied to the raw integrated author database as 38,750 records referred to 32,937 unique authors. For duplicated records, a clerical review was performed (Herzog et al., 2007). Manual verification of duplicate records revealed that 1,207 records (12.15% in terms of duplicated records and 3.11% of all integrated records) were incorrectly assigned to the same person. These records were deleted from the integrated database. An integrated database used in our research finally included 32,937 unique authors of publications, including 25,463 authors of journal articles.

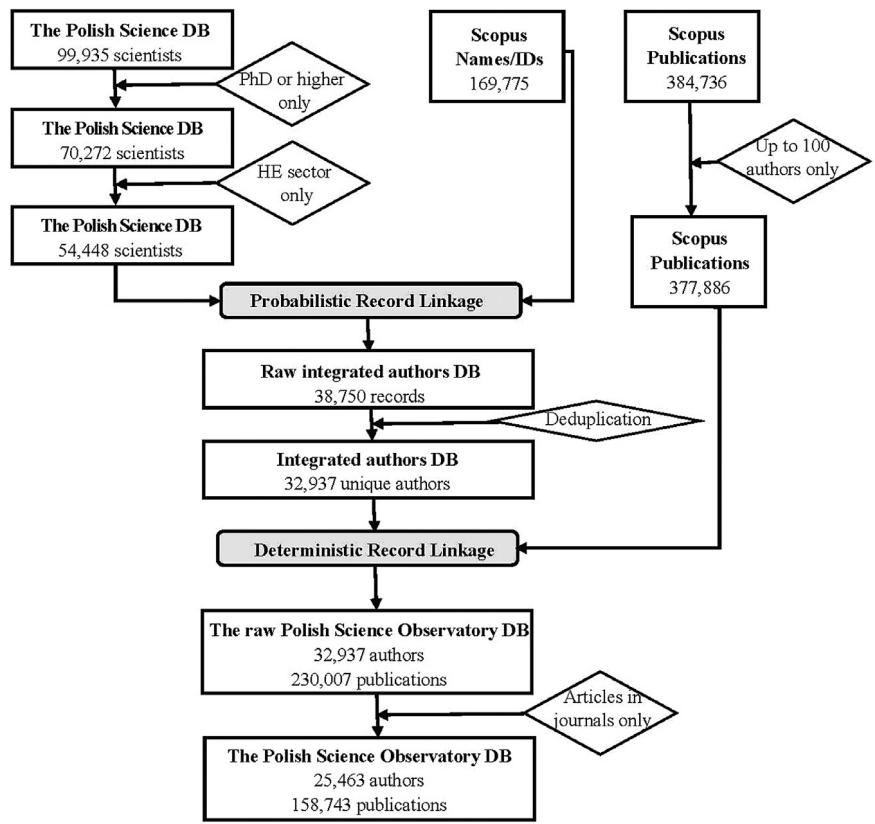

**Fig. 1.** Database merging: Main steps in merging the biographical and administrative dataset (The Polish Science) with the publication and citation database (Scopus).

Finally, Database II also contained metadata on 384,736 publications published in 2009–2018. From among them, the 377,886 papers had up to 100 authors, and 230,007 were written by the authors included in Database I (we used deterministic record linkage at this stage of data integration). Subsequently, only articles written in journals were selected for further analysis, with the number of papers in the database reducing to 158,743 articles.

*3.1.1. Limitations*

Our research has some limitations and possible biases (e.g., selection bias) as a result of the database construction procedures we employ: we select only internationally visible authors, that is, authors with Scopus-indexed publications. The selection of a different database (for instance, Web of Science or the Polish Scientific Bibliography—PBN), a different period (other than 2009–2018), a different publication format (other than articles in journals), or a different language (other than English) might lead to different results.

The date of reference for the data derived from Database I ("The Polish Science") was November 21, 2017, and for the data derived from the Scopus database was the whole decade studied. There are five simplifying assumptions. (1) The paper examines a decade of individual publishing output. While the actual publishing period may, in fact, be shorter than a decade for younger scientists, it may be only the most recent decade of the long-term publishing activities of older scientists. (2) Journal percentile ranks as provided by Scopus are deemed stable within this decade—even though they may fluctuate over the period studied. (3) We assume that Polish scientists were not changing institutions (between 75 research-involved and 10 research-intensive institutions) in the decade studied, as the mobility within the Polish higher education system is very low. (4) We regard Polish scientists who were assistant, associate, and full professors on the date of reference as keeping these positions for the whole decade studied, while these positions are the highest ranks achieved in the study period. (5) We use an internationally recognizable tripartite division of academic positions into assistant, associate, and full professors, even though in fact we use two Polish academic degrees (doctorates and habilitations) and a Polish academic title (professorship). In this sense, our "academic positions" are proxies for Polish "academic degrees and titles". However, all scientists in our sample have their doctoral degrees (and therefore must be at least assistant professors). All scientists with habilitation degrees receive the position of associate professor within three years, and all full professors have their professorship titles.

While biological age, academic position, and employment type and institution were defined as of November 21, 2017, the variables derived from the Scopus database were constructed to show mean values for the decade of 2009–2018, in which they may have differed from year to year. A limitation is that the values for 2017 for some variables and the mean values for the decade of 2009–2018 are lumped together. Clearly, even the binary classification of male-dominated disciplines valid for 2017 may have been different in the previous years, especially for disciplines close to the threshold value of 50% (for instance, HUM, SOC, and ECON, with 49.8%, 49.8%, and 49.1% of female scientists, respectively; see Table 16 in Data Appendices).

This means that longitudinal studies (year by year) and cohort studies (by consecutive cohorts of scientists) were not possible because of data limitations. Actually, the dominant disciplines ascribed to scientists, individual publication portfolios, gender composition of disciplines, and average publication prestige were constructed for the decade of 2009–2018. For instance, a single observation was a male who in 2017 was 60 years old and was employed full-time as an associate professor in a research-intensive university. He was publishing in ASJC Physics and Astronomy, and his individual Scopus-indexed research output for the decade studied (2009–2018) was strictly defined in terms of publication numbers (20 Scopus-indexed journal articles), the gender composition of his co-authorships (60% all-male, 35% mixed-sex, 5% solo), and the individual average publication prestige expressed in percentile ranks (Scopus 85th percentile rank). While acknowledging this limitation, we must stress that Polish scientists publish far too little per year to use Scopus data from a single year (e.g., 2017 only). We assume that a decade of publishing provides a good overview of individual publishing patterns captured within individual publication portfolios.

*3.2. Methods*

As in our previous work on gender disparities in international research collaboration (Kwiek and Roszka, 2020), also here every Polish scientist represented in our integrated database was ascribed to one of 27 ASJC disciplines at the two-digit level (following Abramo, Aksnes, & D'Angelo, 2020, who determined the dominating Web of Science subject category for each scientist they studied). A given paper can have one or multiple disciplinary classifications (see the ASJC discipline codes used, as described in Table 3, which presents the variables used in this analysis). The dominant discipline for each scientist is the mode for each of them: the most frequently occurring value (when no single mode occurred, the dominant discipline was randomly selected). All Polish scientists were defined by their gender, discipline, as well as their publications (solo, all-male, all-female or mixed-sex). Every ASJC discipline represents proportions of male and female scientists. However, GEN, NEURO, and NURS disciplines did not meet an arbitrary minimum threshold of 50 scientists per discipline and were omitted from further analysis.

In the present research, in which the unit of analysis was an individual scientist, every scientist had solo or collaborative articles. Collaborative articles include same-sex and mixed-sex articles. Collaborative articles with authors included in our database are defined in terms of the gender of the authors. Of the Polish scientists included in the integrated database of 54,448 scientists, 100% had their gender defined in the original administrative database. In contrast, there are Polish co-authors outside of our database (e.g., affiliated with the Polish Academy of Sciences) and international co-authors of publications with Polish co-authors whose gender is not defined.

Regarding international collaborators of Polish authors and their gender, we analyzed 158,743 articles with individual EIDs (Scopus individual publication IDs). There were 15,149 articles (9.54%) written solely by female scientists, 39,089 (24.62%) written solely by male scientists, 78,419 (49.40%) written in mixed female-male collaboration, and 18,109 (11.41%) solo-written articles. There were 7,979 articles (5.03%) for which only the gender of Polish co-authors was known.

For the purpose of determining the gender of the international co-authors, we used another dataset at our disposal: a dataset of 27.4 million publications published in the same period of 2009–2018 in the OECD area and indexed in Scopus. Our "OECD" dataset includes all metadata about all publications produced in the study period in 1,674 research-active institutions located in 40 OECD economies (the threshold we used was 3,000 Scopus-indexed articles published in the past 10 years). Specifically, we used a subset of our OECD dataset of authors (with 11,087,392 individual Scopus IDs). In the next step, we used the R package of genderizeR to estimate the gender of the OECD authors from our OECD dataset (see Wais, 2016, on the various gender determination methods, including via the R package).

GenderizeR was previously used for gender prediction in Topaz and Sen (2016) for gender representation in editorial boards in 435 journals in mathematical sciences; Fell and König (2016) studied gender difference in co-authorships among 4,234 industrial-organizational psychologists; Huang et al. (2020) examined gender inequality in the academic careers of 7.9 million Web of Science authors. Finally, Wang et al. (2019) also used the R package to study gender-based homophily in JSTOR publication data. Genderize.io provides a count of the number of times that first name appears in the corpus and corresponding probabilities of gender (which is either male or female). In order to establish optimal values of gender prediction indicators, we can manipulate the threshold of probability and count values.

Using the R package, the gender of 7,640,123 our OECD authors (individual Scopus IDs) was estimated with a probability of greater than or equal to 0.85. With the data at our disposal, out of 11,087,392 authors, the genderizeR algorithm was unable to estimate the gender of 2,521,150 authors (22.74%), including a large number of authors from Japan and South Korea, with whom Poland collaborates only marginally. Out of 8,566,242 authors whose gender the algorithm estimated, in 926,119 (10.81%) of cases, gender was estimated with a probability lower than 0.85. In the next step, using individual Scopus IDs, the "The Polish Science Observatory" and the "OECD" datasets were merged to determine the gender of international collaborators of Polish authors. Out of 164,908 international collaborators, we were able to determine the gender of 83,702 (or 50,75%). Our reference database to estimate the gender of co-authors was restricted to 1,674 research-intensive OECD universities; consequently, we were not able to estimate the gender of collaborators from non-research-intensive universities in the OECD area or from non-OECD universities.

**Table 3**
Variables used in the analysis.

| No. | Variable | Description | Source |
| --- | --- | --- | --- |
| 1. | Biological age | Numerical variable. Biological age as provided by the national registry of scientists (N = 99,935). Age in full years as of 2017 is used. | Observatory |
| 2. | Age group | Categorical variable. Three major age groups are used: young (39 and younger; N = 8,400), middle-aged (40–54; N = 11,014), and older (55 and older; N = 6,049) scientists. | Observatory |
| 3. | Gender | Binary variable, as provided by the national registry of scientists (N = 99,935). No other options are possible in the registry. | Observatory |
| 4. | Academic position | Categorical variable. Three Polish degrees used as proxies of academic positions: doctoral degree only (assistant professor; N = 14,271); habilitation degree (associate professor; N = 7,418); and professorship title (full professor; N = 3,774). All scientists without doctoral degrees and from outside of the higher education sector were removed from the analysis. | Observatory |
| 5. | Discipline | Categorical variable. All scientists ascribed to one of 27 Scopus ASJC disciplines. Dominant disciplines were used (N = 25,463). | Scopus |
| 6. | STEM disciplines | Categorical variable. STEM disciplines: AGRI agricultural and biological sciences; BIO biochemistry, genetics, and molecular Biology; CHEMENG chemical engineering; CHEM chemistry; COMP computer science; DEC decision science; EARTH earth and planetary sciences; ENER energy; ENG engineering; ENVIR environmental science; GEN biochemistry, genetics, and molecular biology; IMMU immunology and microbiology; MATER materials science; MATH mathematics; NEURO neuroscience; NURS nursing; PHARM pharmacology, toxicology, and pharmaceutics; and PHYS physics and astronomy. | Scopus |
| 7. | Non-STEM disciplines | Categorical variable. Non-STEM disciplines: BUS business, management, and accounting; DENT dentistry, ECON economics, econometrics, and finance; HEALTH health professions; HUM arts and humanities; MED medicine; PSYCH psychology; SOC social sciences; and VET veterinary. | Scopus |
| 8. | Male- and female-dominated disciplines | Binary variable. Male-dominated disciplines are those in which the percentage of male scientists exceeds or equals 50% (N = 12,786 scientists). Female-dominated disciplines are those in which the percentage of female scientists exceeds 50% (N = 12,677 scientists). | Observatory |
| 9. | Mean publication prestige (percentile rank) | Categorical variable. Mean prestige represents the median prestige value for all publications written by a scientist in the study period of 2009–2018. For journals for which the Scopus database did not ascribe a percentile rank, we have ascribed the percentile rank of 0; Scopus ascribes percentiles to journals in the 25th to 99th percentile range, with the highest rank being the 99th percentile. | Scopus |
| 10. | Same-sex collaboration ratio | Categorical variable. The percentage of same-sex collaboration articles (male-male, female-female) among all collaborative articles in an individual publication portfolio defined for 2009–2018. | Observatory |
| 11. | Mixed-sex collaboration ratio | Categorical variable. The percentage of mixed-sex collaboration articles (male-female) among all collaborative articles in an individual publication portfolio defined for 2009–2018. | Observatory |
| 12. | Solo collaboration ratio | Categorical variable. The percentage of single-authored articles among all articles in an individual publication portfolio defined for 2009–2018. | Observatory |
| 13. | Research-intensive institution | Binary variable. The 10 institutions (from among 85 examined) are the IDUB (or "Excellence Initiative–Research University") institutions selected in 2019. | Ministry |
| 14. | Number of employees | Numerical variable. The number of full-time employed scientists (in FTEs: full-time equivalents) as of December 31, 2018. | Ministry |

Next, using an individual scientist as the unit of analysis, we calculated the proportion of same-sex publications among collaborative articles within the individual publication portfolio of every Polish scientist in the sample. Thus, for all scientists, male and female, within their collaborative articles only, we determined what we termed the same-sex collaboration ratio (for male scientists collaborating only with male scientists, the ratio is 1). Analogously, the ratio of 0 is equivalent to conducting no same-sex collaboration—the scientist collaborates only with the other gender, i.e., there are only mixed-sex publications in the scientist's individual publication portfolio). The ratio does not take into account the different availability of male and female colleagues within each discipline. The availability, or the gender composition of each discipline, includes both their numbers and their percentages. As the sample section (3.3) shows in detail (see Table 16 in Data Appendices), there are more than 1,000 female scientists in only three disciplines (AGRI, BIO, and MED) and more than 500 in six disciplines (the three above and CHEM, ENG, and ENVIR), whereas there are more than 1,000 male scientists in three disciplines, and more than 500 in 12 disciplines.

The gender composition of the 24 disciplines studied would be a serious limitation of the independent variable as long as we assumed that scientists collaborate only within their disciplines. However, scientists in our dataset collaborate both within disciplines and across them, which can be seen from the disciplinary statistics pertaining to individual publication portfolios by ASJC discipline and to authorship combinations by ASJC discipline for individual papers. Traditionally, especially in academic profession surveys (see Kwiek, 2019), scientists who are strongly embedded in their disciplines are identified, for instance, by checking the discipline of their doctoral dissertations; based on our dataset, in contrast, we examine scientists and their collaborative publications to which disciplines are ascribed via a Scopus indexing system. The availability of male or female colleagues within a discipline seems to matter much less in current settings, given the increasing large-scale cross-disciplinary collaboration (as evidenced by differing dominant ASJC disciplines ascribed to collaborating scientists). Table 3 provides a short description of variables ("Observatory" means The Polish Science Observatory database).

**Table 4**
The median of the same-sex collaboration ratio by gender.

|        | Same-sex collaboration |
|--------|------------------------|
| Male   | 0.500                  |
| Female | 0.153                  |
| Total  | 0.333                  |
| Z      | −44.291                |
| p-value| <0.001                 |

*3.3. Sample*

The characteristics of the sample (N = 25,463; 14,886 males and 10,577 females, 58.5% and 41.5%) is presented in Table 16 in Data Appendices: about half of the scientists are middle-aged (or in the 40-54 age bracket (49.7%), and over half of them are assistant professors (56.0%). Column percentages enable the analysis of the gender distribution by major age groups, academic positions, and disciplines (by type: STEM and non-STEM, female-dominated and male-dominated). Row percentages enable the analysis of how male and female scientists are distributed according to a given feature. About half of the scientists work in female-dominated disciplines and about a half in male-dominated disciplines (49.8% and 50.2%); however, females in female-dominated disciplines are the weaker majority (54.6%) than males in male-dominated disciplines (71.4%). All assistant professors hold doctoral degrees, all associate professors hold habilitations, and all full professors hold professorship titles.

The 25,463 scientists in our integrated database had at least a single article in the Scopus database in the period 2009–2018; therefore, it includes all internationally visible Polish academic scientists (on the skewed distribution of research productivity of Polish scientists, see Kwiek, 2018b; on the upper 10%, termed *top performers*, who produce about half of all publications across 11 European systems, see Kwiek, 2016). Additionally, our sample includes the international collaborators of Polish authors, whose gender was determined using the algorithm described in the Data and Methods subsection (164,908 international co-authors). The differentiated proportions of female scientists can also be examined by academic discipline. While female scientists are especially underrepresented in the four disciplines of computer science (COMP 16.5%), engineering (ENG 14.9%), physics and astronomy (PHYS 16.6%), and mathematics (MATHS 25.2%), the number of male and female scientists is almost equal in arts and humanities (HUM) and social sciences (SOC).

**4. Results**

*4.1. The same-sex collaboration ratio by gender*

**Hypothesis 1.** We would expect that the same-sex collaboration ratio is higher for female than for male scientists (not confirmed).

Gender homophily in publishing, or the same-sex collaboration ratio, falls within the range of 0 (no same-sex collaborative articles among collaborative articles in the individual publication portfolio) to 1 (exclusively same-sex collaborative articles among collaborative articles in the portfolio). The average ratio for males to be involved in same-sex collaboration is more than three times that of females (the median ratio for males is 0.500, compared with 0.153 for females, Table 4). For the whole national sample, the median ratio is 0.333, meaning that at least 50% of authors conduct same-sex collaboration (males with males, females with females) at the 33.3% level. Mann-Whitney's Z-test shows the gender difference to be significantly different at the significance level of 0.05. Thus, Hypothesis 1 is not confirmed.

*4.2. The same-sex collaboration ratio by age and academic position*

**Hypothesis 2.** We would expect that the same-sex collaboration ratio decreases with age for both male and female scientists (confirmed for males but not for females).

**Hypothesis 3.** We would anticipate that the same-sex collaboration ratio decreases with academic position for both male and female scientists (confirmed for males but not for females).

Before analyzing the effect of age and academic position, we examined the level of correlation between these two variables since, in many academic systems, seniority is a significant predictor of career advancement. The boxplots in Fig. 2 divide the data into quartiles and show the median, which is higher for each subsequent academic position. The boxes enclose the middle 50% of the data (for instance, across all disciplines, half of full professors are aged about 60). Outliers are located predominantly above the boxes, showing the presence of older scientists within the three academic positions rather than younger ones. There is a clear interdependence between age and academic position as the average level of age increases with the three consecutive academic positions adopted in this paper (assistant professor, associate professor, and full professor) across all 24 ASJC disciplines. Also, the observed average age for each of the three stages of an academic career is similar among all the disciplines. This empirical observation is confirmed by the formal Kruskal-Wallis test in which we tested the null hypothesis that the average age is the same at every stage of the academic career: for each discipline, we reject the null hypothesis at a significance level of 0.001 (Table 5). However, both variables emerge as important in previous literature and therefore their joint impact will be studied below.

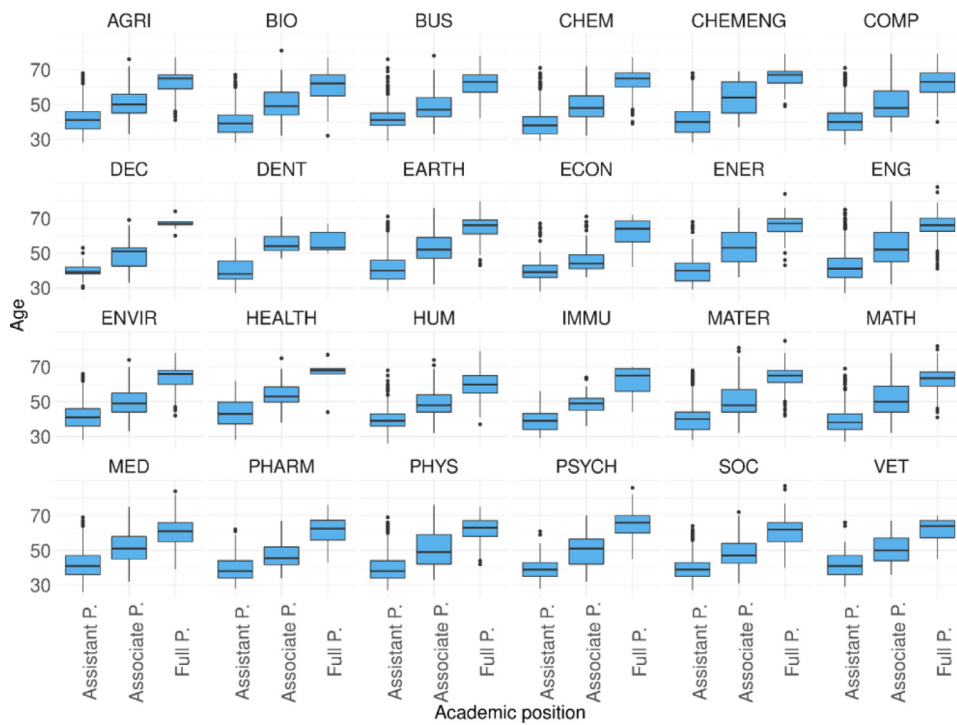

**Fig. 2.** Age distribution in terms of academic position and ASJC.

**Table 5**
Kruskal-Wallis test statistics.

| ASJC | Kruskal-Wallis H | df | Asymp. Sig. | N | ASJC | Kruskal-Wallis H | df | Asymp. Sig. | N |
|---|---|---|---|---|---|---|---|---|---|
| AGRI | 1323.3 | 2 | <0.001 | 2,702 | ENVIR | 668.8 | 2 | <0.001 | 1,680 |
| BIO | 789.9 | 2 | <0.001 | 1,780 | HEALTH | 21.2 | 2 | <0.001 | 67 |
| BUS | 252.7 | 2 | <0.001 | 714 | HUM | 540.9 | 2 | <0.001 | 1,058 |
| CHEM | 744.3 | 2 | <0.001 | 1,475 | IMMU | 64.3 | 2 | <0.001 | 119 |
| CHEMENG | 215.2 | 2 | <0.001 | 481 | MATER | 691.8 | 2 | <0.001 | 1,462 |
| COMP | 370.6 | 2 | <0.001 | 1,030 | MATH | 580.5 | 2 | <0.001 | 1,026 |
| DEC | 31.2 | 2 | <0.001 | 54 | MED | 1498.1 | 2 | <0.001 | 3,574 |
| DENT | 33.4 | 2 | <0.001 | 75 | PHARM | 105.8 | 2 | <0.001 | 254 |
| EARTH | 540.6 | 2 | <0.001 | 1,154 | PHYS | 530.1 | 2 | <0.001 | 1,098 |
| ECON | 131.6 | 2 | <0.001 | 379 | PSYCH | 136.8 | 2 | <0.001 | 304 |
| ENER | 136.2 | 2 | <0.001 | 295 | SOC | 424.9 | 2 | <0.001 | 992 |
| ENG | 1440.9 | 2 | <0.001 | 3,358 | VET | 158.2 | 2 | <0.001 | 332 |

For the purposes of examining the same-sex collaboration ratio by age group, we divided our sample into these three categories: young scientists (aged 39 and younger), middle-aged scientists (aged 40–54) and older scientists (aged 55 and older), of which middle-aged scientists are the largest age group (45.79%) (Table 6). The proportion of males and females is almost equal among young scientists—but females are less than 30% of older scientists (see % column).

Table 7 shows the distribution of the median value of the same-sex collaboration ratio by gender and age group. The median ratio for males slightly decreases with age (by 6 percentage points). In contrast, the same median ratio for females substantially increases with age (by 18 percentage points). While the ratio for females triples with age, it is still very low compared with that of males (the difference being 35 percentage points).

**Table 6**
Distribution of the sample of Polish scientists by age group and gender.

|        |          | Young | Middle-aged | Older | Total |
|--------|----------|-------|-------------|-------|-------|
|        |          | (39 and younger) | (40-54) | (55 and older) | |
| Male   | n        | 3,747 | 6,526 | 4,613 | 14,886 |
|        | % column | 51.2  | 56.0  | 71.2  | 58.5   |
|        | % row    | 25.2  | 43.8  | 31.0  | 100.0  |
| Female | n        | 3,578 | 5,134 | 1,865 | 10,577 |
|        | % column | 48.8  | 44.0  | 28.8  | 41.5   |
|        | % row    | 33.8  | 48.5  | 17.6  | 100.0  |
| Total  | n        | 7,325 | 11,660 | 6,478 | 25,463 |
|        | % column | 100.0 | 100.0 | 100.0 | 100.0  |
|        | % row    | 28.8  | 45.8  | 25.4  | 100.0  |

**Table 7**
The median same-sex collaboration ratio by age group and gender.

|  | Male | Female | Total | Z | p-value |
|---|---|---|---|---|---|
| Young (39 and younger) | 0.5396 | 0.0625 | 0.2727 | -29.676 | <0.001 |
| Middle-aged (40–54)    | 0.5000 | 0.1818 | 0.3333 | -28.163 | <0.001 |
| Older (55 and older)   | 0.4762 | 0.2353 | 0.3750 | -15.696 | <0.001 |
| Total                  | 0.5000 | 0.1538 | 0.3333 | -44.291 | <0.001 |

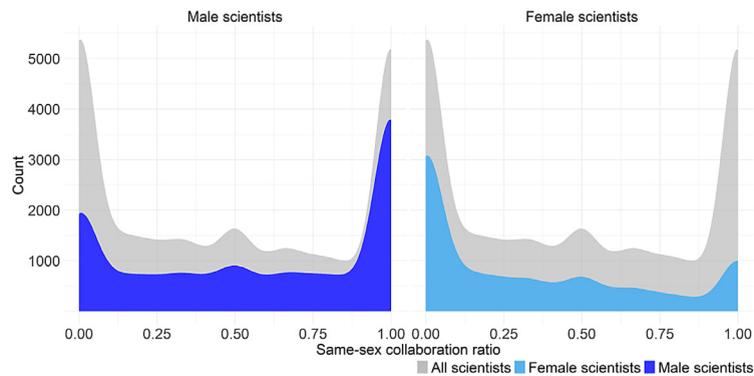

**Fig. 3.** The distribution of the same-sex collaboration ratio by gender. The gray area is the overall distribution for both genders.

The difference in collaboration patterns for young scientists by gender is interesting in view of previous literature about gender patterns of research collaboration. This strand of literature suggests that women tend to co-author with women (Ghiasi et al., 2018; Potthoff & Zimmermann, 2017; Wang et al., 2019; Lerchenmueller et al., 2019), although this is not true in the Polish case. While half of young male scientists write at least 54% of their papers in collaboration with males, the same indicator for females is nine times lower (6.3%). Young males tend to collaborate with males—and young females tend *not* to collaborate with females. While 50% of young female scientists are characterized by the same-sex collaboration ratio at the level of 0.06, in the case of older females, the median ratio quadruples to 0.24: older females still tend to collaborate primarily with males. For all age groups (see the Total line in Table 7), the difference by gender in Polish science is substantial: while the median same-sex collaboration ratio for males is 0.5, the median for females is more than three times lower (0.15). (These results will be confirmed in a fractional logit regression analysis in section 4.6.)

What is clear in the two panels in Fig. 3 is the predominance of extreme values (0 for no same-sex collaboration and 1 for exclusively same-sex collaboration) in individual publication portfolios. The total number of extreme values (0 and 1) is similar for both genders. The majority of collaborations are mixed-sex collaborations. A substantial proportion of collaborating male scientists (left panel, right peak) co-published mostly with males in the decade studied; a substantial proportion of collaborating female scientists (right panel, left peak), in contrast, tended not to co-author with females in the same period.

The distribution of the same-sex collaboration ratio for females is the mirror image of that for males. Apart from the two extreme values of 1 and 0, the distribution of the ratio in question for males is basically uniform. For females, a gradual decline in the ratio is clearly observed. Comparing the extremes, there are more females without same-sex collaboration than males for the same collaboration type; there are about three times more males who collaborate only with males compared with females who collaborate only with females.

When we examine academic positions, in a similar vein, the same-sex collaboration ratio by males decreases with the highest academic position reached (Table 8). In contrast, the same ratio for females increases with academic positions, although its level is still very low for all females. While the median ratio level for females increases two and a half times when we move up the academic ladder, it is still much lower compared with that of males. While 50% of female assistant professors are characterized by the same-sex collaboration ratio at the level of 0.105, for female full professors, the ratio increases to 0.250. See a graphical summary for age groups and academic positions in Fig. 4.

**Table 8**
The median same-sex collaboration ratio by academic position and gender.

|  | Male | Female | Total | Z | p-value |
| --- | --- | --- | --- | --- | --- |
| Assistant Professor | 0.5263 | 0.1053 | 0.3077 | −37.583 | <0.001 |
| Associate Professor | 0.5000 | 0.2083 | 0.3636 | −20.695 | <0.001 |
| Full Professor | 0.3924 | 0.2500 | 0.3333 | −8.840 | <0.001 |
| Total | 0.5000 | 0.1538 | 0.3333 | −44.291 | <0.001 |

The gender difference in collaboration patterns can be studied in more detail using boxplots and violin plots combined. The gender difference by age group (Fig. 5) closely resembles the gender difference by academic position (Fig. 6). Female scientists consistently, across the three age groups and across the three academic positions, tend *not* to collaborate with other females (compare the shapes for Ratio=1, i.e., females collaborating only with females, across the age groups and academic positions for females). Note that the median shown in boxplots is much lower for each group for females than for males, and it increases for females with age; it is also much lower for female assistant and associate professors and lower for female full professors.

Inverse proportionality in collaboration patterns between males and females is visible for each age group and each academic position. In terms of within-sex variation, male scientists are more differentiated than female scientists (compare the height of the boxes in the two columns) for each age group and each academic position studied. Females, and especially young females and female assistant professors, tend not to collaborate with other females. As can be seen from Figs. 5 and 6, generally, conclusions from a study of age groups resemble conclusions from a study of academic positions.

While above, we have studied three broad age groups, below, we focus on biological age as a numerical variable. The year-by-year approach illustrated by regression lines in Fig. 7 generally confirms the two opposite trends for both genders, at least until the age of 60 for males and for all ages for females. Interestingly, the generally downward trend in the same-sex collaboration ratio for male scientists is reversed for those aged 60 and above: the ratio for the oldest males increases. In contrast, for female scientists, the damped growth characteristic of all ages until about 60 turns into exponential growth for the oldest female scientists (a cut-off point of 70 is used, the standard retirement age for full professors). The dots in Fig. 7 represent the median value of the same-sex collaboration ratio for each year of age. Relatively high variation of median values for very young male scientists and no variation for very young female scientists (see the respective dots in both panels) is caused by the low numbers of scientists in these age groups. Thus, Hypotheses 2 and 3 are confirmed for males but not for females.

*4.3. The same-sex collaboration ratio by academic discipline*

**Hypothesis 4.** We would anticipate that the same-sex collaboration ratio is higher in male-dominated academic disciplines (confirmed).

First, we examined the correlation level between the mean same-sex collaboration ratio (ranging from 0 to 1) and the percentage of male scientists within the discipline (see Fig. 8). The correlation between the two variables is weak (r=0.228, $R^2$=0.052); however, as the percentage of males increases, so does the mean same-sex ratio (see the red regression line). The disciplines fall into two categories: female-dominated (left of the vertical dotted line indicating 50%) and male-dominated (right of the line and on the line, by our definition; see Table 1 for the variables). The bubble size reflects the number of scientists. In five disciplines (CHEM, ENVIR, ECON, SOC, and HUM), the percentage of men is very close to 50%. The highest mean same-sex collaboration ratio is not correlated with the male and female distribution within a discipline: it is equally high for physics and astronomy (PHYS) and computer science (COMP), in which male participation exceeds 80%, as it is for pharmacology, toxicology, and pharmaceutics (PHARM) and biochemistry, genetics, and molecular biology (BIO), with male participation in the 30–40% range. At the same time, while social sciences, arts and humanities, and economics, econometrics, and finance (HUM, SOC, and ECON) exhibit a mean same-sex ratio of around 0.5 among the five gender-balanced disciplines (those with close to 50% male participation), chemistry (CHEM) exhibits a ratio of around 0.7.

The same-sex collaboration ratio differs vastly by discipline and by gender. Previous research shows that as the fraction of female researchers in a discipline increases, women increasingly tend to publish with other women; also, the male ratio to co-author with women is higher in disciplines with more women (Boschini & Sjögren, 2007, p. 339). A good way to visualize gender differences in the median same-sex collaboration ratio is through a heat map (the color palette in Table 9 changes from light blue for low values to deep blue for high values). In the case of COMP, ENG, and MATH, with the high overrepresentation of male scientists, the ratio for males is extremely high (and the median values reach the level of 1 or almost 1). That is to say, at least half of male scientists in these disciplines collaborate only with males. In COMP, ENER, ENG, HEALTH, PHYS, and VET, at least half of females do not collaborate with females at all (and the median values reach the level of 0 or almost 0). In contrast, in disciplines such as PHARM, PSYCH, and SOC, the median value for females is significantly higher than for males. The median level by ASJC discipline is also shown graphically in boxplots in Fig. 9. Thus, Hypothesis 4 is confirmed.

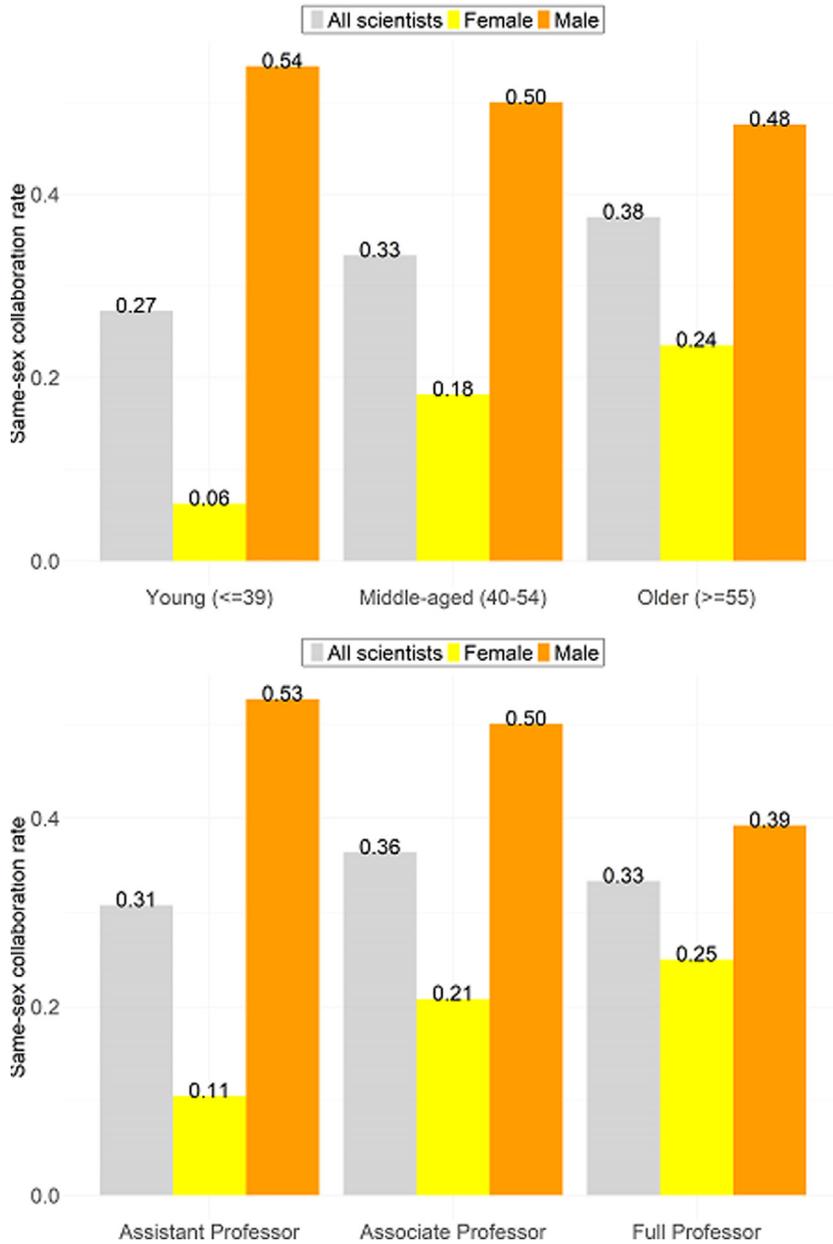

**Fig. 4.** The median same-sex collaboartion ration by gender, age group and academic position.

### 4.4. The same-sex collaboration ratio by institutional type

**Hypothesis 5.** We would expect that the same-sex collaboration ratio is higher in research-intensive universities (confirmed for males but not for females).

Previous literature indicates differences in gender homophily in research collaboration not only by discipline but also by institution. Therefore, we will test whether the same-sex collaboration ratio also differs by institutional type: we contrast the 10 research-intensive institutions with 75 other institutions in the national system. The 10 institutions are the IDUB (or "Excellence Initiative–Research University") institutions, which were selected for additional research funding for the 2020–2026 period. The IDUB institutions include both top Polish universities and polytechnic institutes (similar results were achieved for the top 10 institutions in terms of publication numbers overall and publication numbers within the Scopus 90th–99th journal percentiles).

For male scientists employed in the IDUB institutions, the ratio is high: the proportion of articles published only with males by the upper 50% of male scientists is at least 60% and is larger than the overall ratio for males in the system (see the Total line in

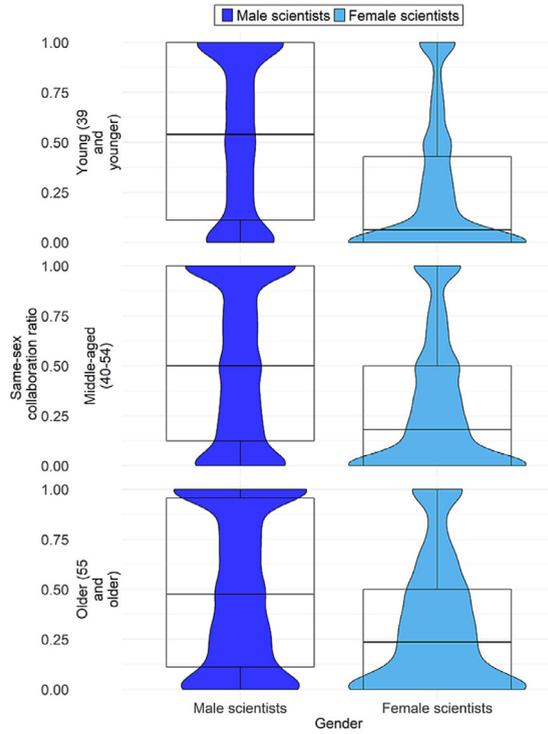

**Fig. 5.** The same-sex collaboration ratio: distribution by age groups and gender (boxplots and violin plots combined).

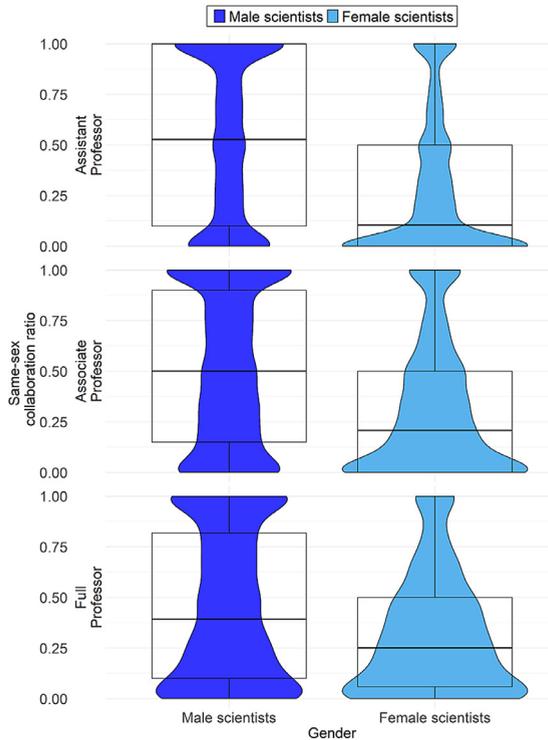

**Fig. 6.** The same-sex collaboration ratio: distribution by academic position and gender (boxplots and violin plots combined).

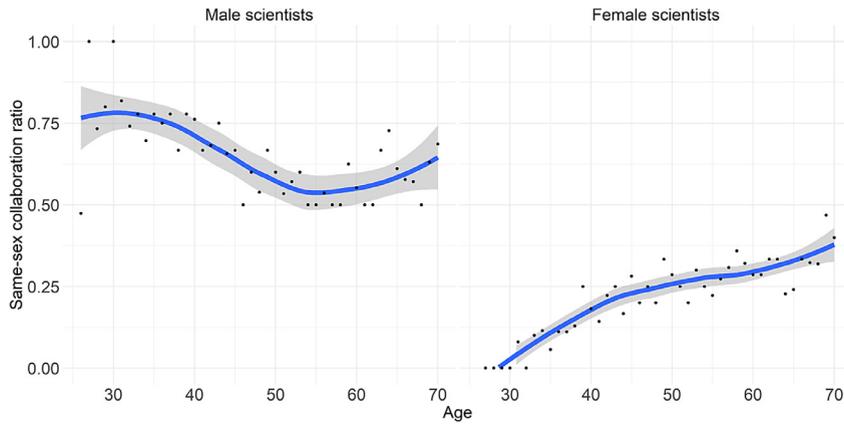

**Fig. 7.** The same-sex collaboration ratio by gender and age. The regression line was estimated using the method of local polynomial regression fitting. The gray area represents 95% confidence intervals. Each year of age is represented by a single dot (a cut-off point of 70 is used). Dots represent median values.

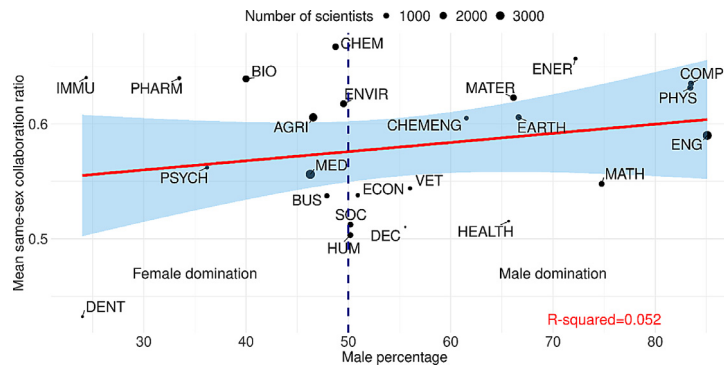

**Fig. 8.** Correlation between mean same-sex ratio and percentage of men in the ASJC disciplines (bubble size reflects the number of scientists).

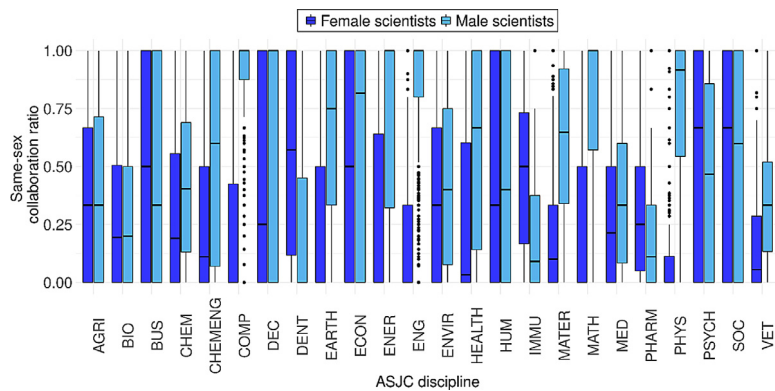

**Fig. 9.** The same-sex collaboration ratio: distribution by discipline and gender.

Table 10: 50%). For female scientists, in contrast, the same proportion in the IDUB institutions is more than four times lower and is even lower than the overall ratio for females in the system. In other words, we reach the somewhat surprising conclusion that for males, the proportion of all-male collaboration in individual publication portfolios is higher in research-intensive institutions than the already high proportion for all institutions—while for females, the proportion of all-female collaboration is lower in research-intensive institutions than the already low proportion for all institutions.

In the Polish academic science system as a whole, the same-sex collaboration ratio for males is more than three times that for females (a finding which is confirmed by a fractional logit regression analysis in Section 4.6 below). Fig. 10 shows the gender difference in the median same-sex collaboration ratio by institutional type and gender in more detail using boxplots and violin plots combined.

Table 9
The median same-sex collaboration ratio by discipline and gender (shading: from the highest ratio in dark blue to the lowest ratio in light blue).

|  | Male | Female | Total | Z | p-value |
|---|---|---|---|---|---|
| AGRI | 0.25 | 0.25 | 0.25 | -0.743 | 0.457 |
| BIO | 0.087 | 0.1111 | 0.1 | -1.384 | 0.166 |
| BUS | 0.5 | 0.5 | 0.5 | -0.49 | 0.624 |
| CHEM | 0.25 | 0.1111 | 0.1818 | -8.753 | <0.001 |
| CHEMENG | 0.4167 | 0.0861 | 0.2566 | -5.403 | <0.001 |
| COMP | 1 | 0 | 0.875 | -13.542 | <0.001 |
| DEC | 1 | 0.75 | 1 | -0.518 | 0.604 |
| DENT | 0 | 0.4118 | 0.2353 | -3.27 | 0.001 |
| EARTH | 0.5714 | 0.1429 | 0.5 | -10.671 | <0.001 |
| ECON | 0.7143 | 0.5 | 0.6667 | -1.456 | 0.145 |
| ENER | 0.75 | 0 | 0.6 | -4.57 | <0.001 |
| ENG | 1 | 0 | 0.8889 | -26.85 | <0.001 |
| ENVIR | 0.2727 | 0.2727 | 0.2727 | -1.31 | 0.19 |
| HEALTH | 0.6333 | 0 | 0.5 | -2.146 | 0.032 |
| HUM | 0.5 | 0.5 | 0.5 | -0.15 | 0.88 |
| IMMU | 0.0909 | 0.3 | 0.25 | -3.033 | 0.002 |
| MATER | 0.5 | 0.1176 | 0.375 | -17.456 | <0.001 |
| MATH | 0.96 | 0.1 | 0.7692 | -16.172 | <0.001 |
| MED | 0.1429 | 0.1148 | 0.125 | -2.113 | 0.035 |
| PHARM | 0.05 | 0.125 | 0.0984 | -3.319 | 0.001 |
| PHYS | 0.6903 | 0 | 0.6 | -16.861 | <0.001 |
| PSYCH | 0.3 | 0.5 | 0.471 | -2.684 | 0.007 |
| SOC | 0.5 | 0.6667 | 0.5 | -2.577 | 0.01 |
| VET | 0.1842 | 0.0455 | 0.1206 | -5.178 | <0.001 |
| Total | 0.5 | 0.1538 | 0.3333 | -44.291 | <0.001 |

Table 10
The median of the same-sex collaboration ratio by institutional type and gender.

| Institutional type | Male | Female | Total | Z | p-value |
|---|---|---|---|---|---|
| Research-intensive (IDUB) | 0.6000 | 0.1348 | 0.4138 | -30.717 | <0.001 |
| Rest | 0.4444 | 0.1667 | 0.2857 | -31.992 | <0.001 |
| Total | 0.5000 | 0.1538 | 0.3333 | -44.291 | <0.001 |

The distribution of the median ratio for females is basically the same in both institutional types, and the within-sex variation is much higher for males than for females, as indicated by the height of the boxplots. The difference between the median values for males and females is much larger in the case of research-intensive institutions; the median value for males is much higher in these institutions, as it is for females.

This effectively means that in research-intensive institutions (see the top IDUB panel in Fig. 10), males as well as females are more likely to collaborate with males. Gender homophily is thus stronger for males and weaker for females in research-intensive institutions. In other institutions (see the bottom panel), the number of males collaborating exclusively with males and the number of males collaborating exclusively with females are equal; the number of females collaborating exclusively with males and the number of females collaborating exclusively with females are similar in both institutional types (see the large base on which the two right columns rest for female scientists in both panels). Thus, Hypothesis 5 is confirmed for males but not confirmed for females.

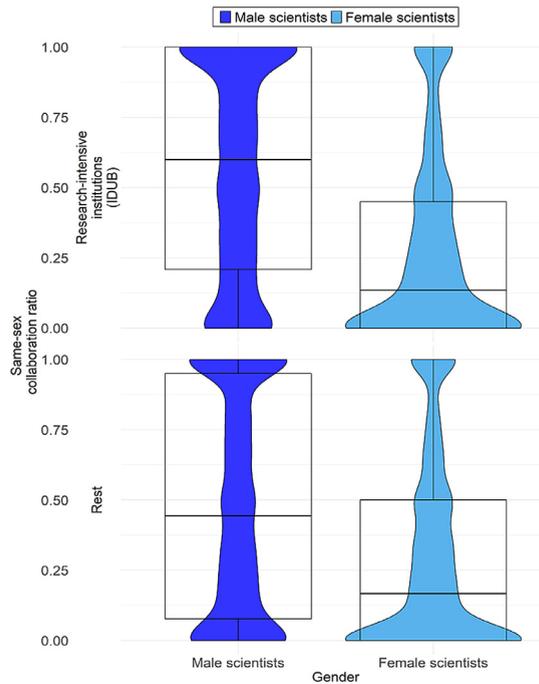

**Fig. 10.** The same-sex collaboration ratio: distribution by institutional type and gender (boxplots and violin plots combined).

**Table 11**
The median prestige level distribution (by percentile from 0–99, with the 99th percentile being the highest) of publications by major gender collaboration type and gender.

|  | Mixed-sex collaboration | Same-sex collaboration | Solo research (zero collaboration) |
| --- | --- | --- | --- |
| Male | 62.50 | 59.17 | 50.00 |
| Female | 62.20 | 58.00 | 46.50 |
| Total | 62.42 | 58.27 | 48.50 |
| Z | -1.497 | -5.981 | -5.121 |
| p-value | 0.134 | <0.001 | <0.001 |

## 4.5. The same-sex collaboration ratio by journal prestige

**Hypothesis 6.** We would expect that the journal prestige level of mixed-sex publications is higher than that of same-sex publications for both male and female scientists (confirmed).

Both the quantity and quality of output in academia are relatively easily measured (with all standard limitations) using the Scopus database as articles are published in journals of different ranks. The scientists in our sample have their own unique individual publication portfolios with publications, translatable into average individual prestige via Scopus citation metrics. The prestige of each article in this portfolio is derived from the prestige of the journal in which it was published and is defined by the percentile rank ascribed annually to each academic journal within its ASJC discipline. Top journals, including the *Journal of Informetrics* and *Scientometrics*, are usually ranked in the upper 5% of journals.

Importantly, the citation-based percentile ranking system used by Scopus is being systematically used in Poland, for instance in a complicated system of indicators used first to select (in 2019) and then to additionally finance (in 2020–2026) 10 research-intensive Polish universities. We used the measure of average prestige, which represents the median prestige value for all publications written by a given scientist in the study period of 2009–2018 for three categories of publications (same-sex, mixed-sex, and solo publications). For journals for which the Scopus database did not ascribe a percentile rank, we have ascribed the percentile rank of 0; Scopus ascribes percentiles to journals in the 25th to 99th percentile range, with the highest rank being the 99th percentile.

The median prestige level (in a range of 0–99) for all Polish publications written in same-sex and mixed-sex collaboration by gender does not differ much (Table 11): the median values for all-male publications and all-female publications by gender are almost identical (59.17 and 58.00, respectively). Also, the median value for mixed-sex collaborations does not differ significantly by gender. Both males and females, on average, regardless of the collaboration type, publish in journals with relatively low prestige. Articles written in mixed-sex collaboration are, on average, published in more prestigious journals than those written in same-sex collaboration and in much more prestigious journals than solo articles (see the Total line in Table 11).

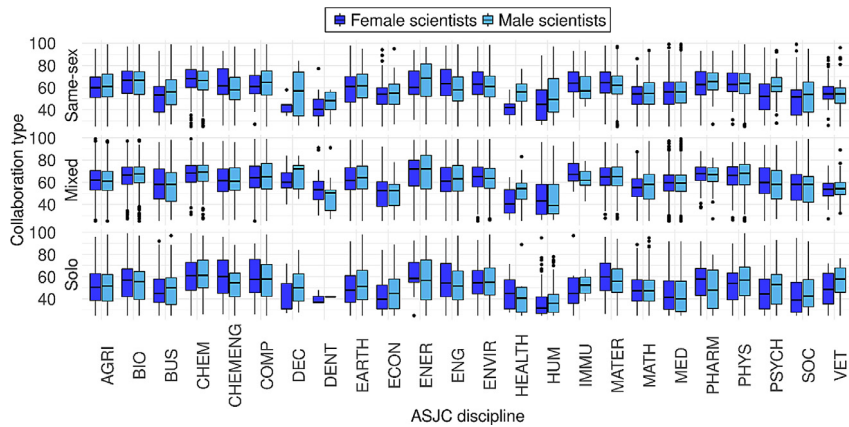

**Fig. 11.** The prestige level distribution of publications (by Scopus percentile rank from 0–99, with the 99$^{th}$ percentile being the highest in prestige) by major collaboration type, gender, and discipline.

The distribution of the median journal prestige level by discipline and collaboration type (mixed, same-sex, and solo publications, separately for males and females) shows both common patterns and substantial variations. Generally, for each ASJC discipline (Table 12), solo research is characterized by the lowest prestige level. BIO, CHEM, ENER, and PHARM belong to disciplines with the highest median prestige level, regardless of the collaboration type. Both mixed-sex and same-sex collaborations have higher average prestige levels than do solo articles.

The differences in prestige level by gender are as follows: for mixed-sex collaborations, they are marginal, but for same-sex collaboration, they are substantial (compare the same-sex collaboration columns for males and females in Table 12). Male-only collaborations have higher median prestige than do female-only collaborations, and this pattern is characteristic of a large number of disciplines. Males collaborating with males, on average, publish in more prestigious journals than do females collaborating with females. Solo research by females exhibits lower median prestige levels than does solo research by males in all except for nine disciplines (including BIO, CHEMENG, ENER, ENG, MATER, MED, and PHARM). The median prestige level by ASJC discipline and gender is also shown graphically in the boxplots in Fig. 11 to go beyond the median values and to highlight intra-disciplinary cross-gender variability, with three separate panels for the three gender-defined collaboration types. Thus, Hypothesis 6 is confirmed.

*4.6. A modeling approach: A fractional logit regression model*

**Hypothesis 7.** In a fractional logit regression model, we would anticipate that individual-level independent variables are more influential than institutional-level independent variables in predicting the same-sex collaboration ratio (not confirmed).

Finally, we move from descriptive statistics and two-dimensional analysis to modeling, and we use a regression model for a fractional dependent variable—a fractional logit regression model (Papke & Woolridge, 1996), designed for variables bounded between zero and one (as with our dependent variable: the same-sex collaboration ratio). Linear models to examine how a set of explanatory variables influences a given proportion or fractional response variable are not appropriate here (Ramalho, Ramalho, & Murteira, 2011, p. 19). In this model, no special data adjustments are needed for the extreme values of zero and one.

In our case, we have 24 ASJC disciplines represented in our 85 research-involved institutions. The number of employees and the percentage of female scientists vary in each of them; each discipline in each institution is either male-dominated (i.e., with exactly or more than 50% male scientists) or female-dominated (i.e., with more than 50% female scientists). We also have a set of 10 highly research-intensive institutions (termed IDUB institutions) and one containing the rest of them. Individual scientists are embedded in their institutions and in their disciplines, and both institutions and disciplines have their specific patterns of cross-gender collaboration. In some disciplines and institutions, same-sex collaboration is more prevalent than in others. For the sake of clarity, here is an example: a single observation here is not a male mathematician with individual features only, such as biological age and academic position. This male mathematician is also embedded in a highly research-intensive institution (variable: IDUB type) employing 2,000 teaching and research faculty (variable: number of employees) and publishing in the discipline of mathematics (variable: STEM discipline), which is male-dominated (variable: male-dominated discipline). Furthermore, in this institution, customarily, male mathematicians tend to have the habit of publishing with their male rather than female colleagues.

In the regression model, we also include the mean individual publication prestige percentile, which requires a clear explanation as it differs from the prestige attached to an individual article (and it may be considered to be a consequence of the collaboration rather than a cause of the collaboration). Collaboration, as defined in this paper, is considered to be a product rather than a process: collaboration between two scientists is viewed only through the proxy of the paper they co-authored and published. By our definition,

**Table 12**

The median prestige level for publications (by Scopus percentile ranks from 0–99, with the 99th percentile being the highest in prestige) by major collaboration type, gender, and discipline (shading: from the highest median prestige level in dark blue to the lowest median prestige level in light blue.

| | Male | | | Female | | | Total | | | Mixed | | Same sex | | Solo | |
|---|---|---|---|---|---|---|---|---|---|---|---|---|---|---|---|
| | Mixed | Same sex | Solo | Mixed | Same sex | Solo | Mixed | Same sex | Solo | Z | p-value | Z | p-value | Z | p-value |
| AGRI | 61 | 59 | 52 | 62 | 59 | 51 | 61 | 59 | 51 | -1.446 | 0.148 | -1.481 | 0.139 | -0.181 | 0.857 |
| BIO | 67 | 65 | 56 | 67 | 63 | 57 | 67 | 64 | 56 | -1.472 | 0.141 | -3.439 | 0.001 | -0.724 | 0.469 |
| CHEM | 68 | 66 | 61 | 68 | 66 | 61 | 68 | 66 | 61 | -0.161 | 0.872 | -0.381 | 0.703 | -0.464 | 0.643 |
| CHEMENG | 59 | 59 | 54 | 62 | 58 | 60 | 60 | 59 | 58 | -1.355 | 0.175 | -0.424 | 0.671 | -1.799 | 0.072 |
| COMP | 67 | 63 | 58 | 65 | 61 | 58 | 67 | 63 | 58 | -1.831 | 0.067 | -0.113 | 0.91 | -0.381 | 0.703 |
| DEC | 75 | 62 | 50 | 67 | 42 | 31 | 74 | 46 | 45 | -0.901 | 0.367 | -2.018 | 0.044 | -1.832 | 0.067 |
| EARTH | 65 | 60 | 51 | 61 | 55 | 48 | 63 | 58 | 50 | -1.846 | 0.065 | -2.727 | 0.006 | -1.943 | 0.052 |
| ENER | 69 | 66 | 57 | 70 | 58 | 59 | 69 | 65 | 58 | -1.084 | 0.278 | -0.713 | 0.476 | -0.662 | 0.508 |
| ENG | 61 | 58 | 52 | 60 | 60 | 54 | 60 | 58 | 52 | -0.534 | 0.594 | -2.067 | 0.039 | -2.867 | 0.004 |
| ENVIR | 63 | 59 | 55 | 64 | 61 | 55 | 64 | 60 | 55 | -2.057 | 0.04 | -2.071 | 0.038 | -0.033 | 0.974 |
| IMMU | 61 | 62 | 53 | 70 | 58 | 45 | 66 | 58 | 45 | -2.695 | 0.007 | -0.728 | 0.467 | -0.471 | 0.637 |
| MATER | 65 | 61 | 56 | 65 | 63 | 60 | 65 | 62 | 58 | -0.278 | 0.781 | -1.293 | 0.196 | -1.799 | 0.072 |
| MATH | 58 | 55 | 47 | 54 | 53 | 48 | 56 | 54 | 47 | -2.446 | 0.014 | -1.411 | 0.158 | -0.103 | 0.918 |
| PHARM | 67 | 62 | 48 | 67 | 63 | 58 | 67 | 63 | 57 | -0.702 | 0.483 | -0.053 | 0.957 | -0.298 | 0.765 |
| PHYS | 68 | 64 | 57 | 66 | 61 | 54 | 67 | 63 | 57 | -1.461 | 0.144 | -1.58 | 0.114 | -1.529 | 0.126 |
| BUS | 58 | 56 | 50 | 58 | 54 | 45 | 58 | 55 | 47 | -0.312 | 0.755 | -2.198 | 0.028 | -1.15 | 0.25 |
| DENT | 51 | 48 | 42 | 47 | 38 | 37 | 48 | 40 | 40 | -0.037 | 0.97 | -1.03 | 0.303 | -0.447 | 0.655 |
| ECON | 54 | 55 | 45 | 50 | 48 | 40 | 53 | 52 | 42 | -0.498 | 0.618 | -1.337 | 0.181 | -1.863 | 0.062 |
| HEALTH | 54 | 55 | 41 | 45 | 44 | 45 | 50 | 48 | 45 | -1.517 | 0.129 | -2.668 | 0.008 | -0.365 | 0.715 |
| HUM | 46 | 51 | 36 | 43 | 45 | 32 | 45 | 48 | 33 | -0.992 | 0.321 | -1.796 | 0.072 | -2.157 | 0.031 |
| MED | 59 | 56 | 40 | 59 | 53 | 42 | 59 | 54 | 41 | -1.392 | 0.164 | -4.63 | 0 | -0.383 | 0.702 |
| PSYCH | 64 | 59 | 53 | 60 | 49 | 45 | 61 | 54 | 48 | -0.679 | 0.497 | -3.151 | 0.002 | -0.985 | 0.325 |
| SOC | 58 | 54 | 43 | 58 | 53 | 39 | 58 | 53 | 41 | -0.304 | 0.761 | -1.32 | 0.187 | -1.876 | 0.061 |
| VET | 54 | 54 | 58 | 54 | 54 | 49 | 54 | 54 | 52 | -0.558 | 0.577 | -1.034 | 0.301 | -0.906 | 0.365 |
| Total | 63 | 59 | 50 | 62 | 58 | 47 | 62 | 58 | 49 | -1.446 | 0.148 | -1.481 | 0.139 | -0.181 | 0.857 |

**Table 13**

Fractional logistic regression model statistics, dependent variable: the same-sex collaboration ratio (N = 21,467).

| $R^2 = 0.159$ | Estimate | Robust std. errors | t value | Pr(>|t|) | VIF |
|---|---|---|---|---|---|
| (Constant) | -0.873 | 0.096 | -9.087 | <0.001 | |
| Age | 0.005 | 0.001 | 3.497 | <0.001 | 2.030 |
| Male | 0.890 | 0.024 | 36.760 | <0.001 | 1.147 |
| IDUB | 0.214 | 0.031 | 6.851 | <0.001 | 1.823 |
| Full Professor | -0.250 | 0.043 | -5.820 | <0.001 | 2.061 |
| Associate Professor | -0.060 | 0.029 | -2.102 | 0.036 | 1.391 |
| STEM discipline | -0.174 | 0.055 | -3.136 | 0.002 | 1.037 |
| Mean prestige points | -0.002 | 0.001 | -2.260 | 0.024 | 1.078 |
| Number of employees | -0.00004 | 0.00002 | -2.509 | 0.012 | 1.798 |
| Male-dominated discipline | 0.744 | 0.023 | 32.050 | <0.001 | 1.133 |

every scientist in our dataset has his or her own, clearly defined mean individual publication prestige percentile, which is determined by the entirety of their Scopus-indexed publication output from the decade studied (each article is linked to its source or journal, with a clear Scopus-calculated highest journal percentile). Consequently, the mean individual publication prestige percentile for each scientist (with the 99[th] percentile being the highest) is an individual-level predictor: it is a proxy for the average prestige of their maximally decade-long publication history. It is higher for scientists publishing exclusively in top journals and lower for those publishing in a combination of top and second-tier journals or in second-tier journals only, as ranked by Scopus.

As our dependent variable is fractional (ranging from zero to one), we estimate a fractional logit regression model. We estimate odds ratios for conducting same-sex collaboration in journal publishing, i.e., publishing with scientists of the same sex. We calculate the same-sex collaboration ratio as the percentage of same-sex collaboration articles in all of the published collaborative articles in all of the scientists' individual publication portfolios. Using a fractional logistic regression approach, we estimated the probability of conducting same-sex collaboration.

The first type of independent variable captures scientists' individual demographic, biographical, and bibliometric characteristics: gender, biological age, mean individual publication prestige level within the study period of 10 years (or less), current academic position, and the type of the dominant Scopus-defined ASJC discipline (STEM or non-STEM). The second type of independent variable captures three major institutional characteristics: a binary variable indicating employment in an IDUB or non-IDUB institution (being employed full-time in one of the 10 highly research-intensive institutions or not), the number of scientists employed in the author's institution (in FTEs in 2018), and publishing in a male-dominated discipline or not. The nonexistence of collinearity of the independent variables was confirmed through an analysis of VIF coefficients (see Table 17 in Data Appendices). Although the correlation table of independent variables shows in some cases (e.g. IDUB – the number of employees, full professor – age) a pairwise correlation of moderate strength, the vector of independent variables is not characterized by significant collinearity, as indicated by the VIF coefficients (Table 13). The correlation between these pairs is largely controlled by other variables in the model.

The distribution of residuals in our dataset was not normal (i.e., the K-S normality test statistic is equal to 0.104, with a p-value less than 0.001). The normality of residuals distribution allows performing statistical inference on the model properties as all statistical significance tests assume the normality of distribution. To overcome the model's inconsistency with the assumptions, robust standard errors were estimated, and, on the basis of the estimates, a significance test for individual coefficients in the model was conducted. A further step in the analysis of the residuals distribution indicates the lack of influential observations (as the range of standardized residuals does not exceed ± 3 standard deviations). Consequently, the conclusions drawn from our model are valid.

Table 13 presents the results of a fractional logit model. All coefficients are statistically significant and have a significant impact on the same-sex collaboration ratio. However, the parameters of the model thus estimated cannot be interpreted naturally. As Long and Freese (2006) show, for a natural and transparent interpretation of the influence of independent variables on the same-sex collaboration ratio, average marginal effects should be estimated for the model (see Table 15). The greatest influence on the ratio is exerted by gender and publishing in male-dominated disciplines. Being a male scientist increases the ratio on average by 0.2 (which means that male scientists on average have a 20 percentage points higher same-sex collaboration ratio than female scientists, *ceteris paribus*). Publishing in a male-dominated discipline increases this ratio on average by 0.167. Working in a highly research-intensive (IDUB type) institution increases the ratio on average by 0.048, and the influence of age is relatively weak: every ten years of age results in an average increase of the ratio by 0.01. Being a full professor decreases on average the ratio by 5.6 p.p., and being an associate professor decreases it by 1.4 p.p. Also publishing mainly in STEM disciplines on average decreases the ratio by 3.9 p.p. The weakest, though still significant, predictors of the same-sex collaboration ratio are the journal prestige and the number of employees in scientists' institution.

Thus, Hypothesis 7 is not confirmed: the impact on gender homophily is exerted by both individual-level and institutional-level predictors, rather than by merely individual-level ones. Being male is as influential in the model as working in a male-dominated discipline and being employed in a research-intensive university. There are two most influential predictors of the same-sex collaboration ratio, both gender-related: being male and working in a male-dominated discipline.

As $R^2$ is relatively low ($R^2 = 0.159$), there are other, unknown predictors of conducting same-sex collaboration in academic publishing. Other predictors are known predominantly from qualitative literature about research collaboration (see, especially, Sonnert

& Holton, 1996; Lerchenmueller et al., 2019); however, the data necessary to be used as explanatory variables in our model are not available. Other predictors determining gender homophily in research and in gender network structures more generally include individual, disciplinary, institutional, cultural, political, social, and economic factors, as previous national qualitative and quantitative studies indicate (see, e.g., McPherson et al., 2001; McDowell & Smith, 1992, who included 178 US male and female economists with their full publication records; Ibarra, 1992, who conducted 79 interviews with employees of an advertising firm; Ibarra, 1997, who conducted 63 interviews with middle-level US managers; Kegen, 2013, who studied two German Excellence Initiative institutions; and Feenay & Bernal, 2010, who conducted a survey of scientists and engineers from 151 US research universities). Especially useful seem to be the various studies from systems known for their high female participation in academic science (Halevi, 2019; see, especially, the cross-national data in Elsevier, 2020; Elsevier 2018).

## 5. Summary of findings, discussion, and conclusions

Our research differs from previous studies in several respects. First, we examined every internationally publishing Polish male and female scientist and the entirety of internationally visible (Scopus-indexed) Polish academic knowledge production for a decade (2009–2018). Second, owing to the characteristics of the database used, we had 100% gender determination for all scientists in the system (rather than probability thresholds in gender determination). Third, we defined what we termed the "individual publication portfolio" for every Polish scientist to examine the same-sex collaboration ratio at the level of the individual scientist (the idea of the individual publication portfolio can be used for other research questions: see our research on gender disparities in international research collaboration—Kwiek & Roszka, 2020). Fourth, our unit of analysis was the gender-defined individual scientist rather than the individual publication, with his or her specific distribution of male/female authorships.

Finally, and most importantly, we used a comprehensive, fully integrated biographical, administrative, publication, and citation database (the "The Polish Science Observatory" database, which we constructed by merging the national registry of all 99,535 Polish scientists with the Scopus dataset comprised of all their publications in 2009–2018). Our sample (N = 25,463) included all the university professors holding at least a doctoral degree and employed in 85 research-involved universities, grouped into 27 disciplines with all their Scopus-indexed publications (158,743 articles).

While most previous literature highlights that women are much more likely to have a female than a male co-author (in three top economic journals, Boschini & Sjögren, 2007; in life sciences, Holman & Morandin, 2019; among computer scientists, Jadidi et al., 2018; and among industrial-organizational psychologists, Fell & König, 2016), or a female rather than a male collaborator in research projects (Lerchenmueller et al., 2019), leading to excessive gender homophily in female publishing, our findings, which are based on a large national sample, do not support this gender disparity in collaboration patterns. These findings may tend to indicate the uniqueness of massively expanded and relatively gender-balanced Central European science systems, with female participation in the academic labor force reaching 50%, testifying to the significance of proportions, or minorities and majorities, for academic life (Kanter, 1977).

Having an integrated dataset at our disposal, we were able to examine the same-sex collaboration ratio across several dimensions, previously usually either studied separately and with only a selection of our variables or studied based on small datasets. This research goes beyond traditional bibliometric studies of gender-based homophily in research collaboration by combining the following:

(1) biographical and administrative data routinely inaccessible to large-scale studies, namely, the biological age of all scientists (rather than a proxy of first publication) and the three stages of their academic careers (assistant, associate, and full professorships), as well as
(2) data that is routinely accessible in bibliometric studies, such as journal prestige, academic disciplines, and institutional type.

Previous research tended to be restricted either 1) by focusing on selected institutions (Kegen, 2013) or selected disciplines (McDowell & Smith, 1992; Lerchenmueller et al., 2019; Fell & König, 2016; Maddi et al., 2019), sometimes with disciplines represented by their top journals (Potthoff & Zimmermann, 2017; Boschini & Sjögren, 2007), or 2) by being large in scale but focused solely on bibliometric data (Huang et al., 2020; Wang et al., 2019; Ghiasi et al., 2015; Larivière et al., 2013; Ghiasi et al., 2018). This research, in contrast, reveals the opportunities that large-scale, comprehensive national databases may provide (such databases are currently available for Norway and Italy; see Abramo, Aksnes, & D'Angelo, 2020, who compared the performance of Norwegian and Italian professors and verified the feasibility of applying their "research efficiency indicator" to the two systems). Although our "Observatory" database is not an example of the Current Research Information System (CRIS) as a data source as recently defined by Sivertsen (2019), new Polish databases (such as POL-on 2.0 and PBN, which are national registries of all higher education institutions, publications, and scientists) are moving in the CRIS direction.

Our results show that in the Polish academic science system as a whole, the same-sex collaboration ratio for males is more than three times that for females (a finding which is confirmed by a fractional logit regression analysis: it is on average 20 percentage points higher for males). The ratio for females to collaborate with females and for males to collaborate with males (or gender homophily in publishing patterns) showed clear patterns in accordance with biological age and academic seniority: across all age groups, female scientists tend to collaborate with male scientists, and male scientists also tend to collaborate with male scientists. All-female collaboration, often discussed in literature (Boschini & Sjögren, 2007; McDowell & Smith, 1992; McDowell, Singell, & Stater, 2006), is marginal, and all-male collaboration is pervasive. The gender patterns in publishing are stable not only across age groups—but also across academic positions. Both males and females, on average, regardless of the gender-defined collaboration type (same-sex, mixed-sex, or solo publications), publish in journals with prestige that is relatively low. However, articles written in mixed-sex collaboration are, on average, published in more prestigious journals than those written in same-sex collaboration (which

**Table 14**
Residuals statistics of the fractional logit regression model.

|  | Standard residual |
|---|---|
| Mean | 0.000 |
| Median | -0.007 |
| Std. Deviation | 1.000 |
| Range | 4.109 |
| Minimum | -2.023 |
| Maximum | 2.086 |
| K-S normality test statistic | 0.104 |
| p-value | <0.001 |

**Table 15**
Fractional logistic regression model statistics, average marginal effects (N = 21,467).

| $R^2 = 0.159$ | Estimate | Robust st. errors | t value | Pr(>|t|) |
|---|---|---|---|---|
| Age | 0.001 | 0.000 | 3.499 | <0.001 |
| Male | 0.200 | 0.005 | 39.853 | <0.001 |
| IDUB | 0.048 | 0.007 | 6.865 | <0.001 |
| Full Professor | -0.056 | 0.010 | -5.827 | <0.001 |
| Associate Professor | -0.014 | 0.006 | -2.102 | 0.036 |
| STEM discipline | -0.039 | 0.012 | -3.138 | 0.002 |
| Mean prestige points | -0.0004 | 0.0002 | -2.260 | 0.024 |
| Number of employees | -0.000010 | 0.000004 | -2.510 | 0.012 |
| Male-dominated discipline | 0.167 | 0.005 | 33.952 | <0.001 |

is consistent with previous literature; see Campbell, Mehtani, Dozier, & Rinehart, 2013; see the prestige economy of top journals in (Kwiek, 2021)

The difference in collaboration patterns for young scientists (an age group with equal participation of males and females in our sample) by gender is especially interesting in view of previous literature about gender patterns of research collaboration. Previous research suggests that women tend to co-author with women (Ghiasi et al., 2018; Potthoff & Zimmermann, 2017; Wang et al., 2019; Lerchenmueller et al., 2019)—which is not the Polish case. While half of young male scientists write at least 54% of their papers in collaboration with males, the same indicator for females (writing with females) is nine times lower (6.3%). So, while young males tend to collaborate with males, young females tend *not* to collaborate with females. While 50% of young female scientists are characterized by the same-sex collaboration ratio at the level of 0.06, in the case of older females, the median ratio quadruples to 0.24: older females still tend to collaborate primarily with males. For all age groups, the difference by gender in Polish science is substantial: while the median same-sex collaboration ratio for males is 0.5, the median for females is only 0.15.

This finding is not in line with previous research, which generally shows that female scientists exhibit stronger gender homophily than male scientists (Jadidi et al., 2018) and that females collaborate more often with females than males with males (Kegen, 2013; Ghiasi et al., 2018). Gender homophily in team formation (Boschini & Sjögren, 2007) in Poland seems to occur with male scientists but not with female scientists. One explanation might be along the lines of the gender and competition theme introduced in Section 2: younger females, feeling more "under the magnifying glass" and being less "aggressive, combative, and self-promoting" in seeking higher visibility (Sonnert & Holton, 1996, pp. 67–69), tend to co-author with males rather than with other females because males are viewed as more deeply embedded in science. Also academic norms may be viewed as influencing publishing patterns, including, for instance, predominantly same-sex publishing for young male scientists and predominantly mixed-sex publishing for young female scientists (considering that the availability of male and female colleagues under 40 in the system is similar). To verify this hypothesis further, we would need the biological age and academic position of all of the international co-authors, not only the Polish ones, which could be represented only via different proxies.

The gender difference in the same-sex collaboration ratio by age group closely resembles the gender difference by academic position. Male scientists in general tend to collaborate with males; female scientists consistently, across the three age groups and across the three academic positions, tend not to collaborate with other females. Inverse proportionality in collaboration between the two genders is characteristic of each age group and each academic position.

The year-by-year approach we used generally confirms the two opposite trends for both genders: the downward trend in the same-sex collaboration ratio for male scientists stands in sharp contrast to the upward trend for female scientists. In the specific Polish case, age and academic positions are strongly correlated as the principle of "up or out" has not been operative in the system for at least three decades.

We have examined the same-sex collaboration ratio across all disciplines. Differently than in most previous studies, we compared male-dominated disciplines with female-dominated disciplines. Our research supports the finding from previous research that as the fraction of female researchers in a discipline increases, females increasingly tend to co-author with other females (Boschini & Sjö-gren, 2007). In the case of the male-dominated fields of computer science, engineering, and mathematics, the same-sex collaboration ratio for males is prodigious: at least half of male scientists in these disciplines collaborate exclusively with males. In computer science, engineering, health professions, and physics and astronomy, at least half of females do not collaborate with females at all. In contrast, in several female-dominated disciplines (e.g., social sciences and psychology), the median value of same-sex collaboration for females is significantly higher than for males.

The same-sex collaboration ratio also differs by institutional type. We contrasted 10 highly research-intensive institutions with 75 other institutions. An interesting conclusion is that for males, the proportion of all-male collaboration in individual publication portfolios is higher in research-intensive institutions than the already high proportion for all institutions—while for females, the proportion of all-female collaboration is lower in research-intensive institutions than the already low proportion for all institutions. Males in research-intensive institutions are even more likely to collaborate with males, and females are even less likely to collaborate with females. Gender homophily in research-intensive institutions is thus stronger for males and weaker for females than in the rest of the higher education system, which might suggest that a stronger institutional research focus generally induces collaboration with male scientists.

Finally, using a fractional logistic regression approach, we estimated the strength and direction of predictors of conducting same-sex collaboration. The model showed that the same-sex collaboration ratio for male scientists is on average 20 percentage points higher than that for female scientists and that working in a male-dominated discipline increases the ratio by 16.7 p.p.; age slightly increases the ratio (which is in line with our findings from two-dimensional analyses for females but not for males). Also being employed in a highly research-intensive institution increases the ratio by 4.8 p.p. Finally, individual-level factors do not emerge from this model as more influential than institutional-level factors: both types of factors matter. Being male and working in a male-dominated discipline are the two most influential predictors of the same-sex collaboration ratio, followed by working in a research-intensive university.

Male-female collaboration practices in research were tested against the homophily principle: our findings indicate that similarity indeed breeds connection between individual scientists and structures academic publishing ties. However, in the Polish case, this is true only for male scientists. Gender-based homophily has substantial implications for academic careers, with the citation measure being increasingly used as a "reward currency in science" that often underlies decisions on all major aspects of an academic career (Ghiasi et al., 2018, p. 1519). While forming collaborative research teams—perhaps more intuitively and resulting from the dominant social norms in academia rather than from solid individual publishing strategies—Polish female scientists tend not to publish with other females and seem to prefer male co-authors (perhaps viewing male scientists as hubs in colaboative science attracting more attention, with power and resources such as e.g. the supervision of doctoral students). This is the case not only for young female scientists (who may have had predominantly male mentors in their doctoral programs), but for all age groups. This, in time, paradoxically, may contribute to the reduction of the gender productivity, citation, and promotion gaps in Polish science as, conversely, previous global literature suggests that these gaps may widen if females excessively co-author and form professional networks with females, especially in male-dominated disciplines (Maliniak et al., 2013). However, more detailed studies, especially those based on surveys and interviews, would be needed to exclude the possibility that female scientists feel coerced to add male co-authors as part of the global problem of authorship manipulation in academic research (Fong & Wilhite, 2017).

Future research avenues would include, first, moving to a global study: from our "The Polish Science Observatory" dataset to our parallel "The OECD Science Observatory" dataset (with a complete set of metadata pertaining to all 19.3 million articles produced in the same study period of 2009–2018 in 1,674 research-active institutions located in 40 OECD economies, with their 8.7 million unique authors). A global account would allow examining gender-based homophily from a comparative cross-national perspective. Second, future research should also include moving from a cross-sectional study in which "individual publication portfolios" come from a single decade to a longitudinal study in which the portfolios come, for instance, from the 1990s, 2000s, and 2010s, and are compared, thereby revealing cross-national and global trends in man-woman collaboration over time and from a more historical perspective.

**Author contributions**

Marek Kwiek: Conceived and designed the analysis; Collected the data; Contributed data or analysis tools; Performed the analysis; Wrote the paper.

Wojciech Roszka: Conceived and designed the analysis; Contributed data or analysis tools; Performed the analysis; Wrote the paper.

**Acknowledgements**

Acknowledgments The authors gratefully acknowledge the support of the Ministry of Science and Higher Education through its Dialog grant 0022/DLG/2019/10 (RESEARCH UNIVERSITIES).

**Data Appendices**

Tables 16 **and** 17

**References**

Abramo, G., D'Angelo, C. A., & Rosati, F (2015). Selection committees for academic recruitment: Does gender matter? *Res. Eval., 24*(4), 392–404.

**Table 16**

Structure of the sample, all Polish internationally visible university professors, by gender, age group, academic position, and discipline (by type: ASJC, STEM and non-STEM, female-dominated and male-dominated), presented with column and row percentages (Young scientists mean those 39 and younger, middle-aged those 40-54, and older those aged 55 and more).

|   |   | Female | | | Male | | | Total | | |
|---|---|---|---|---|---|---|---|---|---|---|
|   |   | N | % col | % row | N | % col | % row | N | % col | % row |
| Age group | 26 - 30 | 245 | 2.3 | 46.8 | 279 | 1.9 | 53.2 | 524 | 2.1 | 100.0 |
|   | 31 - 35 | 1,653 | 15.6 | 49.8 | 1,664 | 11.2 | 50.2 | 3,317 | 13.0 | 100.0 |
|   | 36 - 40 | 2,148 | 20.3 | 47.1 | 2,411 | 16.2 | 52.9 | 4,559 | 17.9 | 100.0 |
|   | 41 - 45 | 2,272 | 21.5 | 45.7 | 2,696 | 18.1 | 54.3 | 4,968 | 19.5 | 100.0 |
|   | 46 - 50 | 1,569 | 14.8 | 43.8 | 2,013 | 13.5 | 56.2 | 3,582 | 14.1 | 100.0 |
|   | 51 - 55 | 993 | 9.4 | 40.3 | 1,471 | 9.9 | 59.7 | 2,464 | 9.7 | 100.0 |
|   | 56 - 60 | 671 | 6.3 | 37.7 | 1,108 | 7.4 | 62.3 | 1,779 | 7.0 | 100.0 |
|   | 61 - 65 | 563 | 5.3 | 26.7 | 1,548 | 10.4 | 73.3 | 2,111 | 8.3 | 100.0 |
|   | 66 - 70 | 417 | 3.9 | 23.0 | 1,396 | 9.4 | 77.0 | 1,813 | 7.1 | 100.0 |
|   | 71+ | 46 | 0.4 | 13.3 | 300 | 2.0 | 86.7 | 346 | 1.4 | 100.0 |
|   | Total | 10,577 | 100.0 | 41.5 | 14,886 | 100.0 | 58.5 | 25,463 | 100.0 | 100.0 |
| Academic position | Assistant Prof. | 6,851 | 64.8 | 48.0 | 7,420 | 49.8 | 52.0 | 14,271 | 56.0 | 100.0 |
|   | Asssoc. Prof. | 2,822 | 26.7 | 38.0 | 4,596 | 30.9 | 62.0 | 7,418 | 29.1 | 100.0 |
|   | Full Professor | 904 | 8.5 | 24.0 | 2,870 | 19.3 | 76.0 | 3,774 | 14.8 | 100.0 |
|   | Total | 10,577 | 100.0 | 41.5 | 1,886 | 100.0 | 58.5 | 25,463 | 100.0 | 100.0 |
| Discipline (ASJC) – STEM | AGRI | 1,444 | 13.7 | 53.4 | 1,258 | 8.5 | 46.6 | 2,702 | 10.6 | 100.0 |
|   | BIO | 1,068 | 10.1 | 60.0 | 712 | 4.8 | 40.0 | 1,780 | 7.0 | 100.0 |
|   | CHEM | 756 | 7.1 | 51.3 | 719 | 4.8 | 48.7 | 1,475 | 5.8 | 100.0 |
|   | CHEMENG | 185 | 1.7 | 38.5 | 296 | 2.0 | 61.5 | 481 | 1.9 | 100.0 |
|   | COMP | 170 | 1.6 | 16.5 | 860 | 5.8 | 83.5 | 1,030 | 4.0 | 100.0 |
|   | DEC | 24 | 0.2 | 44.4 | 30 | 0.2 | 55.6 | 54 | 0.2 | 100.0 |
|   | EARTH | 385 | 3.6 | 33.4 | 769 | 5.2 | 66.6 | 1,154 | 4.5 | 100.0 |
|   | ENER | 82 | 0.8 | 27.8 | 213 | 1.4 | 72.2 | 295 | 1.2 | 100.0 |
|   | ENG | 501 | 4.7 | 14.9 | 2 857 | 19.2 | 85.1 | 3 358 | 13.2 | 100.0 |
|   | ENVIR | 848 | 8.0 | 50.5 | 832 | 5.6 | 49.5 | 1 680 | 6.6 | 100.0 |
|   | IMMU | 90 | 0.9 | 75.6 | 29 | 0.2 | 24.4 | 119 | 0.5 | 100.0 |
|   | MATER | 495 | 4.7 | 33.9 | 967 | 6.5 | 66.1 | 1 462 | 5.7 | 100.0 |
|   | MATH | 259 | 2.4 | 25.2 | 767 | 5.2 | 74.8 | 1 026 | 4.0 | 100.0 |
|   | PHARM | 169 | 1.6 | 66.5 | 85 | 0.6 | 33.5 | 254 | 1.0 | 100.0 |
|   | PHYS | 182 | 1.7 | 16.6 | 916 | 6.2 | 83.4 | 1 098 | 4.3 | 100.0 |
| Discipline (ASJC) – non-STEM | BUS | 372 | 3.5 | 52.1 | 342 | 2.3 | 47.9 | 714 | 2.8 | 100.0 |
|   | DENT | 57 | 0.5 | 76.0 | 18 | 0.1 | 24.0 | 75 | 0.3 | 100.0 |
|   | ECON | 186 | 1.8 | 49.1 | 193 | 1.3 | 50.9 | 379 | 1.5 | 100.0 |
|   | HEALTH | 23 | 0.2 | 34.3 | 44 | 0.3 | 65.7 | 67 | 0.3 | 100.0 |
|   | HUM | 527 | 5.0 | 49.8 | 531 | 3.6 | 50.2 | 1 058 | 4.2 | 100.0 |
|   | MED | 1 920 | 18.2 | 53.7 | 1 654 | 11.1 | 46.3 | 3 574 | 14.0 | 100.0 |
|   | PSYCH | 194 | 1.8 | 63.8 | 110 | 0.7 | 36.2 | 304 | 1.2 | 100.0 |
|   | SOC | 494 | 4.7 | 49.8 | 498 | 3.3 | 50.2 | 992 | 3.9 | 100.0 |
|   | VET | 146 | 1.4 | 44.0 | 186 | 1.2 | 56.0 | 332 | 1.3 | 100.0 |
|   | Total | 10 577 | 100.0 | 41.5 | 14 886 | 100.0 | 58.5 | 25 463 | 100.0 | 100.0 |
| Gender domination in disciple. | Female-dom. | 6 918 | 65.4 | 54.6 | 5 759 | 38.7 | 45.4 | 12677 | 49.8 | 100.0 |
|   | Male-dom. | 3 659 | 34.6 | 28.6 | 9 127 | 61.3 | 71.4 | 12 786 | 50.2 | 100.0 |
|   | Total | 10 577 | 100.0 | 41.5 | 14 886 | 100.0 | 58.5 | 25 463 | 100.0 | 100.0 |
| Mean publication prestige (percentile) | <0,30) | 777 | 7.3 | 48.7 | 817 | 5.5 | 51.3 | 1 594 | 6.3 | 100.0 |
|   | <30,40) | 888 | 8.4 | 41.3 | 1262 | 8.5 | 58.7 | 2 150 | 8.4 | 100.0 |
|   | <40,50) | 1432 | 13.5 | 39.7 | 2171 | 14.6 | 60.3 | 3 603 | 14.1 | 100.0 |
|   | <50,60) | 2778 | 26.3 | 40.8 | 4023 | 27.0 | 59.2 | 6 801 | 26.7 | 100.0 |
|   | <60,70) | 2573 | 24.3 | 40.8 | 3728 | 25.0 | 59.2 | 6 301 | 24.7 | 100.0 |
|   | <70,80) | 1691 | 16.0 | 43.4 | 2202 | 14.8 | 56.6 | 3 893 | 15.3 | 100.0 |
|   | <80,90) | 373 | 3.5 | 39.4 | 573 | 3.8 | 60.6 | 946 | 3.7 | 100.0 |
|   | <90,100) | 65 | 0.6 | 37.1 | 110 | 0.7 | 62.9 | 175 | 0.7 | 100.0 |
|   | Total | 10 577 | 100.0 | 41.5 | 14 886 | 100.0 | 58.5 | 25 463 | 100.0 | 100.0 |

Abramo, G., Aksnes, D. W., & D'Angelo, C. A (2020). Comparison of research productivity of Italian and Norwegian professors and universities. *J. Informetr., 14*(2), Article 101023.

Aksnes, D. W., Rørstad, K., Piro, F. N., & Sivertsen, G. (2011). Are female researchers less cited? A large scale study of Norwegian researchers. *J. Am. Soc. Inf. Sci. Tech., 62*(4), 628–636.

Aksnes, D. W., Piro, F. N., & Rørstad, K. (2019). Gender gaps in international research collaboration: A bibliometric approach. *Scientometrics, 120*, 747–774.

Antonowicz, D., Kulczycki, E., & Budzanowska, A. (2020). Breaking the deadlock of mistrust? A participative model of the structural reforms in higher education in Poland. *High Educ. (Dordr)* on-line first February 14, 2020. 10.1111/hequ.12254.

Bieliński, J., & Tomczyńska, A. (2018). The ethos of science in contemporary Poland. *Minerva, 57*(2), 151–173.

Boschini, A., & Sjögren, A. (2007). Is team formation gender neutral? Evidence from coauthorship patterns. *J. Labor Econ., 25*(2), 325–365.

Campbell, L. G., Mehtani, S., Dozier, M. E., & Rinehart, J. (2013). Gender-heterogeneous working groups produce higher quality science. *PLOS ONE, 8*(10), e79147.

Dargnies, M.-P. (2012). Men too sometimes shy away from competition: The case of team competition. *Manag. Sci., 58*(11), 1982–2000.

Diezmann, C., & Grieshaber, S. (2019). *Women professors. Who makes it and how?*. Singapore: Springer Nature.

**Table 17**

Inverted matrix of correlation between the predictors.

| | Age | Male | IDUB (research-intensive) | Full Professor | Associate Professor | STEM field | Mean prestige (percentile) | Number of employees | Male-dominated field |
|---|---|---|---|---|---|---|---|---|---|
| Age | 1.000 | 0.169 | -0.027 | 0.588 | 0.206 | 0.044 | -0.120 | -0.028 | 0.018 |
| Male | 0.169 | 1.000 | 0.094 | 0.153 | 0.043 | 0.071 | 0.016 | 0.034 | 0.290 |
| IDUB (research-intensive) | -0.027 | 0.094 | 1.000 | 0.020 | 0.009 | 0.000 | 0.145 | 0.658 | 0.170 |
| Full Professor | 0.588 | 0.153 | 0.020 | 1.000 | -0.266 | 0.035 | 0.003 | 0.028 | 0.007 |
| Associate Professor | 0.206 | 0.043 | 0.009 | -0.266 | 1.000 | -0.021 | 0.001 | -0.001 | -0.005 |
| STEM field | 0.044 | 0.071 | 0.000 | 0.035 | -0.021 | 1.000 | 0.114 | -0.047 | -0.071 |
| Mean prestige (percentile) | -0.120 | 0.016 | 0.145 | 0.003 | 0.001 | 0.114 | 1.000 | 0.169 | -0.016 |
| Number of employees | -0.028 | 0.034 | 0.658 | 0.028 | -0.001 | -0.047 | 0.169 | 1.000 | 0.082 |
| Male-dominated field | 0.018 | 0.290 | 0.170 | 0.007 | -0.005 | -0.071 | -0.016 | 0.082 | 1.000 |

Elsevier. (2020). *The researcher journey through a gender lens*. Amsterdam: Elsevier.
Elsevier. (2018). *Gender in the global research landscape*. Amsterdam: Elsevier.
Enamorado, T., Fifield, B., & Imai, K. (2019). Using a probabilistic model to assist merging of large-scale administrative records. *Am. Political Sci. Rev., 113*(2), 353–371.
Feeney, M. K., & Bernal, M. (2010). Women in STEM networks: Who seeks advice and support from women scientists? *Scientometrics, 85*(3), 767–790.
Feldy, M., & Kowalczyk, B. (2020). The ethos of science and the perception of the Polish system of financing science. *Eur. Rev., 28*(4), 599–616.
Fell, C. B., & König, C. J. (2016). Is there a gender difference in scientific collaboration? A scientometric examination of co-authorships among industrial-organizational psychologists. *Scientometrics, 108*(1), 113–141.
Fellegi, I. P., & Sunter, A. B. (1969). A theory for record linkage. *J. Am. Stat. Assoc., 64*, 1183–1210.
Flory, J. A., Leibbrandt, A., & List, J. A. (2014). Do competitive workplaces deter female workers? A large-scale natural field experiment on job entry decisions. *Rev. Econ. Stud., 82*(1), 122–155.
Fong, E. A., & Wilhite, A. W. (2017). Authorship and citation manipulation in academic research. *PLoS ONE, 12*(12), Article e0187394.
Fox, M. F. (2020). Gender, science, and academic rank: Key issues and approaches. *Quantitative Science Studies, 1*(3), 1001–1006.
Ghiasi, G, Mongeon, P., Sugimoto, C., & Larivière, V. (2018). Gender homophily in citations. In *3rd International Conference on Science and Technology Indicators (STI 2018)* (pp. 1519–1525).
Ghiasi, G., Larivière, V., & Sugimoto, C. R. (2015). On the compliance of women engineers with a gendered scientific system. *PLOS ONE*, (12), 10.
GUS. (2019). *Higher education institutions and their finances in 2018*. Warsaw: Statistics Poland.
Halevi, G. (2019). Bibliometric studies on gender disparities in science. In W. Glänzel, H. F. Moed, U. Schmoch, & M. Thelwall (Eds.), *Springer handbook of science and technology indicators* (pp. 563–580). Cham: Springer.
Harron, K., Dibben, C., Boyd, J., Hjern, A., Azimaee, M., Barreto, M. L., & Goldstein, H. (2017). *Challenges in administrative data linkage for research*. Big Data Soc July–December, 1–12.
Herzog, T. N., Scheuren, F. J., & Winkler, W. E. (2007). *Data quality and record linkage techniques*. Dordrecht: Springer.
Holman, L., & Morandin, C. (2019). Researchers collaborate with same-gendered colleagues more often than expected across the life sciences. *PLOS ONE, 14*(4), Article e0216128.
Huang, J., Gates, A. J., Sinatra, R., & Barabási, A.-L. (2020). Historical comparison of gender inequality in scientific careers across countries and disciplines. *Proc. Natl. Acad. Sci. U.S.A., 117*(9), 4609–4616.
Ibarra, H. (1992). Homophily and differential returns: Sex differences in network structure and access in an advertising firm. *Adm. Sci. Q., 37*(3), 422–447.
Ibarra, H. (1997). Paving an alternative route: Gender differences in managerial networks. *Soc. Psychol. Q., 60*(1), 91.
Jadidi, M., Karimi, F., Lietz, H., & Wagner, C. (2018). Gender disparities in science? Dropout, productivity, collaborations, and success of male and female computer scientists. *Adv. Complex Syst., 21*(3–4), Article 1750011.
Jaro, M. A. (1989). Advances in record linkage methodology as applied to the 1985 census of Tampa Florida. *J. Am. Stat. Assoc., 84*(406), 414–420.
Kanter, R. M. (1977). Some Effects of Proportions on Group Life: Skewed Sex Ratios and Responses to Token Women. *American Journal of Sociology, 82*(5), 965–990.
Kegen, N. V. (2013). Science networks in cutting-edge research institutions: Gender homophily and embeddedness in formal and informal networks. *Procedia Soc. Behav. Sci., 79*, 62–81.
King, M. M., Bergstrom, C. T., Correll, S. J., Jacquet, J., & West, J. D. (2017). Men set their own cites high: Gender and self-citation across fields and over time. *Socius, 3*.
Kulczycki, E., Korzeń, M., & Korytkowski, P. (2017). Toward an excellence-based research funding system: Evidence from Poland. *J. Informetr, 11*(1), 282–298.
Kulczycki, E., & Korytkowski, P. (2020). Researchers publishing monographs are more productive and more local-oriented. *Scientometrics, 125*, 1371–1387.
Kosmulski, M. (2015). Gender disparity in Polish science by year (1975–2014) and by discipline. *J. Informetr., 9*(3), 658–666.
Kwiek, M. (2012). Changing higher education policies: From the deinstitutionalization to the reinstitutionalization of the research mission in Polish universities. *Sci. Public Policy, 35*(5), 641–654.
Kwiek, M. (2016). The European research elite: A cross-national study of highly productive academics across 11 European systems. *High Educ. (Dordr), 71*(3), 379–397.
Kwiek, M. (2018a). Academic top earners. Research productivity, prestige generation and salary patterns in European universities. *Sci. Public Policy, 45*(1, February), 1–13.
Kwiek, M. (2018b). High research productivity in vertically undifferentiated higher education systems: Who are the top performers? *Scientometrics, 115*(1), 415–462.
Kwiek, M. (2019). *Changing European academics. A comparative study of social stratification, work patterns and research productivity*. London and New York: Routledge.
Kwiek, M. (2020a). What large-scale publication and citation data tell us about international research collaboration in Europe: Changing national patterns in global contexts. *Stud. High. Educ., 45*, 1–21 On-line first April 10, 2020. 10.1080/03075079.2020.1749254.
Kwiek, M. (2020b). Internationalists and locals: International research collaboration in a resource-poor system. *Scientometrics, 125*, 57–105.
Kwiek, M., & Roszka, W. (2020). Gender disparities in international research collaboration: A large-scale bibliometric study of 25,000 university professors. *J. Econ. Surv.* Online first 13 November 2020. 10.1111/joes.12395.
Kwiek, M. (2021). The Prestige Economy of Higher Education Journals: A Quantitative Approach. *High Educ. (Dordr), 81*, 493–519.
Larivière, V., Vignola-Gagné, E., Villeneuve, C., Gelinas, P., & Gingras, Y. (2011). Sex differences in research funding, productivity and impact: An analysis of Quebec university professors. *Scientometrics, 87*(3), 483–498.
Larivière, V., Sugimoto, C. R., Chaoquin, N., Gingras, Y., & Cronin, B. (2013). Global gender disparities in science. *Nature, 504*, 211–213.

Lerchenmueller, M., Hoisl, K., & Schmallenbach, L. (2019). *Homophily, biased attention, and the gender gap in science. Paper presented at DRUID19*. Copenhagen Business School June 19-21, 2019.
Long, J. S., & Freese, J. (2006). *Regression models for categorical dependent variables using Stata*. College Station, Texas: Stata Press.
Maddi, A., Larivière, V., & Gingras, Y. (2019). Man-woman collaboration behaviors and scientific visibility: Does gender affect the academic impact in economics and management? In *Proceedings of the 17th International Conference on Scientometrics & Informetrics* (pp. 1687–1697). September 2–5, 2019.
Maliniak, D., Powers, R., & Walter, B. F. (2013). The gender citation gap in international relations. *Int. Organ., 67*(4), 889–922.
McDowell, J. M., Larry, D., Singell, Jr., & Stater, M. (2006). Two to tango? Gender differences in the decisions to publish and coauthor. *Econ. Inq., 44*(1), 153–168.
McDowell, J. M., & Smith, J. K. (1992). The effect of gender-sorting on propensity to coauthor: Implications for academic promotion. *Econ. Inq., 30*(1), 68–82.
McPherson, M., Smith-Lovin, L., & Cool, J. M. (2001). Birds of a feather: Homophily in social networks. *Annu. Rev. Sociol., 27*, 415–444.
Mihaljević-Brandt, H., Santamaría, L., & Tullney, M. (2016). The effect of gender in the publication patterns in mathematics. *PLOS ONE, 11*(10), Article e0165367.
Niederle, M., & Vesterlund, L. (2007). Do women shy away from competition? Do men compete too much? *Q. J. Econ., 122*(3), 1067–1101. 10.1162/qjec.122.3.1067.
Nielsen, M. W. (2016). Gender inequality and research performance: Moving beyond individual-meritocratic explanations of academic advancement. *Stud. High. Educ., 41*(11), 2044–2060.
Papke, L. E., & Wooldridge, J. M. (1996). Econometric methods for fractional response variables with an application to 401(k) plan participation rates. *J. Appl. Econ., 11*, 619–632.
Potthoff, M., & Zimmermann, F. (2017). Is there a gender-based fragmentation of communication science? An investigation of the reasons for the apparent gender homophily in citations. *Scientometrics, 112*(2), 1047–1063.
Ramalho, E. A., Ramalho, J. J. S., & Murteira, J. M. R (2011). Alternative estimating and testing empirical strategies for fractional regression models. *J. Econ. Surv., 25*(1), 19–68.
Mayer, S. J., & Rathmann, J. M. K (2018). How does research productivity relate to gender? Analyzing gender differences for multiple publication dimensions. *Scientometrics, 117*(3), 1663–1693.
Sarsons, H., Gërxhani, K., Reuben, E., & Schram, A. (2021). Gender differences in recognition for group work. *J. Political Econ., 129*(1), 101–147.
Shaw, M. A. (2019). Strategic instrument or social institution: Rationalized myths of the university in stakeholder perceptions of higher education reform in Poland. *Int. J. Educ. Dev., 69*, 9–21.
Siemienska, R. (2007). The puzzle of gender research productivity in Polish universities. In R. Siemienska, & A. Zimmer (Eds.), *Gendered career trajectories in academia in cross-national perspectives* (pp. 241–266). Warsaw: Scholar.
Sivertsen, G. (2019). Developing Current Research Information Systems (CRIS) as data sources for studies of research. In W. Glänzel, H. F. Moed, U. Schmoch, & M. Thelwall (Eds.), *Springer handbook of science and technology indicators* (pp. 667–683). Cham: Springer.
Sonnert, G., & Holton, G. (1996). Career patterns of women and men in the sciences. *Am. Sci., 84*(1), 63–71 JSTOR.
Topaz, C. M., & Sen, S. (2016). Gender representation on journal editorial boards in the mathematical sciences. *PLOS ONE, 11*(8), Article e0161357.
Van den Besselaar, P., & Sandström, U. (2016). Gender differences in research performance and its impact on careers: A longitudinal case study. *Scientometrics, 106*(1), 143–162.
Van den Besselaar, P., & Sandström, U. (2015). Early career grants, performance, and careers: A study on predictive validity of grant decisions. *J. Informetr., 9*(4), 826–838.
Wagner, C. S. (2018). *The Collaborative Era in Science. Governing the Network*. Cham: Palgrave Macmillan.
Wang, Y. S., Lee, C. J., West, J. D., Bergstrom, C. T., & Erosheva, E. A. (2019). Gender-based homophily in collaborations across a heterogeneous scholarly landscape. ArXiv:1909.01284 [Stat]. http://arxiv.org/abs/1909.01284
Winkler, W. (1990). String comparator metrics and enhanced decision rules in the Fellegi-Sunter model of record linkage. In *Proceedings of the Section on Survey Research Methods* (pp. 354–359). American Statistical Association.
Wuchty, S., Jones, B. F., & Uzzi, B. (2007). The increasing dominance of teams in production of knowledge. *Science, 316*(5827), 1036–1039.
Xie, Y., & Shauman, K. A. (2003). *Women in science. Career processes and outcomes*. Cambridge, MA: Harvard University Press.
Zippel, K. (2017). *Women in global science*. Stanford: Stanford University Press.

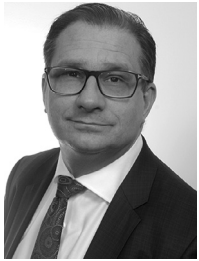

**Marek Kwiek**, Professor (full) and Director, Institute for Advanced Studies in Social Sciences and Humanities (IAS), University of Poznan, Poland. His research area is quantitative studies of science, sociology of science and higher education research. His focus is on global science, international research collaboration, academic productivity, and global academic elites. His recent monograph is *Changing European Academics. A Comparative Study of Social Stratification, Work Patterns and Research Productivity* (Routledge 2019) and he authored 210 papers (*Scientometrics, Science and Public Policy, Journal of Economic Surveys, Higher Education, Studies in Higher Education* etc.). A Principal Investigator or country Team Leader in 25 international research projects. An international expert for the European Commission, USAID, OECD, World Bank, UNESCO, Council of Europe, and the European Parliament. He spent three years at North American universities, including the University of Virginia, UC Berkeley, and McGill University. An editorial board member for *Higher Education Quarterly*, *European Educational Research Journal*, *British Educational Research Journal*, and *European Journal of Higher Education.*.

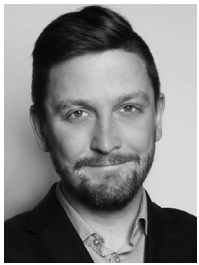

**Dr. Wojciech Roszka** is an assistant professor in the Department of Statistics, the Institute of Informatics and Quantitative Economics, Poznan University of Economics and Business. His area of research interest and expertise includes integration of data from different sources with particular emphasis on probabilistic record linkage, statistical matching, data fusion and microsimulation modelling. He uses these methods primarily in the analysis of the higher education market, the labour market and demographics. He is also involved in projects related to marketing research, time series forecasting and methods of multivariate statistics. He is a member of the Polish Statistical Society and works in the Center for Public Policy Studies of the University of Poznan.